\documentclass[preprint,pre,eqsecnum,longbibliography]{revtex4-1} 

\usepackage{natbib}
\usepackage{graphicx}
\usepackage{amssymb}
\usepackage{amsfonts}
\usepackage{amsmath}
\usepackage{amsbsy}
\usepackage{mathrsfs} 
\usepackage{color}
\usepackage{gensymb} 
\usepackage{hyperref} 
\usepackage{soul}
\usepackage{lineno}
\usepackage{esint}
\usepackage{mymacros}

\begin{document} 

\title{Leidenfrost levitation of a spherical particle above a liquid bath: evolution of the vapour-film morphology with particle size}
\author{Rodolfo Brand{\~a}o}
\author{Ory Schnitzer}
\affiliation{Department of Mathematics, Imperial College London, London SW7 2AZ, UK}

\begin{abstract}
We consider a spherical particle levitating above a liquid bath owing to the Leidenfrost effect, where the vapour of either the bath or sphere forms an insulating film whose pressure supports
the sphere's weight. Starting from a reduced formulation based on a lubrication-type approximation, we use matched asymptotics to describe the morphology of the vapour film assuming that the sphere is small relative to the capillary length (small Bond number) and that the densities of the bath and sphere are comparable.
We find that this regime is comprised of two formally infinite sequences of distinguished limits which meet at an accumulation point, the limits being defined by the smallness of an intrinsic evaporation number relative to the Bond number. These sequences of limits reveal a surprisingly intricate evolution of the film morphology with increasing sphere size. 
Initially, the vapour film transitions from a featureless morphology, where the thickness profile is parabolic, to a neck-bubble morphology, which consists of a uniform-pressure bubble bounded by a narrow and much thinner annular neck. Gravity effects then become important in the bubble leading to sequential formation of increasingly smaller neck-bubble pairs near the symmetry axis. This process terminates when 
the pairs closest to the symmetry axis become indistinguishable and merge.
Subsequently, the inner section of that merger transitions into a uniform-thickness 
film that expands radially, gradually squishing larger and larger neck-bubble pairs into a region of localised oscillations sandwiched between the uniform film and what remains of the bubble whose radial extent is presently comparable to the uniform film; the neck-bubble pairs farther from the axis remain essentially intact. Ultimately, the uniform film gobbles up the largest outermost bubble, whereby the morphology  simplifies to a uniform film bounded by localised oscillations. 
Overall, the asymptotic analysis describes the continuous evolution of the vapour film
from a neck-bubble morphology typical of a Leidenfrost drop levitating above a flat solid substrate to a uniform-film morphology which resembles that in the case of a large liquid drop levitating above a liquid bath. 
\end{abstract}
\maketitle

\section{Introduction}\label{sec:intro}
In the classical Leidenfrost effect, a liquid drop levitates above a superheated solid substrate on a cushion of its own vapour \cite{Biance:03}. The levitation is associated with a sharp transition from nucleate to film boiling above a critical substrate temperature, which is termed the `Leidenfrost point'. The temperature of the drop is approximately constant and equal to the liquid's boiling temperature, which is normally significantly lower than the Leidenfrost point. In this setup, the levitation is enabled by a lubrication pressure field associated with the  flow of vapour escaping from beneath the drop. Most often the evaporation is weak in the sense that the thickness of the vapour layer, on the order of tens of microns, is much smaller than its millimetre-scale radius.

While the vapour film is typically flat relative to its horizontal extent, it may nevertheless exhibit nontrivial morphology relative to its thickness. Such morphology has been the subject of many experimental and theoretical studies, especially owing to foreseen  connections with various dynamical phenomena and instabilities associated with Leidenfrost drops \cite{Quere:13}. Thus, Celestini \textit{et al.}~\cite{Celestini:12} and Burton \textit{et al.}~\cite{Burton:12} used interferometry to observe that the vapour film is composed of a vapour bubble region bounded by a narrow and much thinner annular neck region. Shortly following these observations, Sobac \textit{et al.}~\cite{Sobac:14} put forward a model where a lubrication approximation of the vapour film is `patched' at some radius to a hydrostatic description of the top side of the drop. The model was solved numerically, showing excellent agreement with the experiments for a wide range of drop sizes. An asymptotic analysis of this model, in the limit where an intrinsic evaporation number is small, was also presented in that paper. While the asymptotic results are numerically not very accurate, the analysis provided an insightful linkage between the observed neck-bubble morphology and the weak-evaporation limit. 

Pomeau \textit{et al.} \cite{Pomeau:12} formulated a simpler lubrication model specifically suitable to a drop which is small compared to the capillary length. In that case, the drop shape is approximately a sphere except near the small `flat spot' at the  bottom  of the drop. This simpler model also predicts the formation of a neck-bubble morphology as a suitably defined evaporation number is decreased, or, equivalently, as drop size is increased, starting from an essentially featureless morphology where film thickness varies quadratically with radial distance from the symmetry axis. Unfortunately, the asymptotic analysis in \cite{Pomeau:12} of the corresponding weak-evaporation limit was shown to be erroneous \cite{Sobac:14}. 

Recently there is interest in variations on the classical Leidenfrost effect where the solid substrate is replaced by a liquid substrate. This has several advantages, such as the temperature difference required for levitation being relatively low and the absence of the `chimney' instability which in the classical scenario leads to the breakup of large Leidenfrost drops. In particular, an evaporating liquid drop can be made to levitate above a heated liquid bath \cite{Maquet:16}. Additionally, an inert liquid drop, or solid particle for that matter, can be made to levitate above an evaporating cryogenic liquid bath; this is called the inverse-Leidenfrost effect \cite{Hall:69,Adda:16}.

The morphology of the vapour film in configurations where the substrate is a deformable liquid bath can be quite different than in the classical Leidenfrost effect involving a flat solid substrate. This was demonstrated by Maquet \textit{et al}.~\cite{Maquet:16}, who adapted the lubrication approach of Sobac \textit{et al}.~\cite{Sobac:14} to model their experiments of ethanol drops levitated above a heated silicone-oil bath. Numerical solutions of the adapted model showed that, as  drop size is increased, the film morphology evolves from one that is analogous to the classical Leidenfrost effect (i.e., a featureless parabolic thickness profile followed by formation of a neck-bubble structure) to one where a film of seemingly 
uniform thickness is bounded by localised oscillations.

The latter scenario featuring a seemingly uniform film 
was elucidated by van Limbeek \textit{et al}.~\cite{Van:19} through asymptotic analysis in a weak-evaporation limit, 
specifically for drops large relative to the capillary length and having the same density as the liquid bath. Their analysis provided an analytical description of the film region observed in \cite{Maquet:16}, which was shown to asymptotically correspond to a region where pressure variations are controlled by gravitational-hydrostatic contributions; in that description, the film thickness in fact varies to leading order, although numerically the variation is small.
Furthermore, the oscillations at the edge of that film were identified to be essentially the same as the gravity-capillary waves first studied by Jones and Wilson \cite{Jones:78,Wilson:83}, 
which have since been encountered in various guises by many authors  \cite{Bowles:95,Jensen:97,Duchemin:05,Benilov:08,Benilov:10,Cuesta:12,Cuesta:14,Benilov:15,Hewitt:15}. Asymptotically, such capillary waves are characterised by a localised, formally infinite, chain of  regions. 

As explained by van Limbeek \textit{et al}.~\cite{Van:19}, their large-drop asymptotic analysis breaks down for sufficiently small drops, essentially because the capillary waves ultimately penetrate into the seemingly uniform film. That analysis, therefore, does not asymptotically describe the overall evolution of the 
morphology with increasing drop size as observed numerically in \cite{Maquet:16}. 

Our goal here is to illuminate that evolution by considering a closely related yet simpler 
physical setting, where a spherical particle levitates above a liquid bath owing to the Leidenfrost or inverse-Leidenfrost effect. To that end, we shall employ scaling arguments and the method of matched asymptotic expansions \cite{Hinch:book} to describe the double limit where evaporation is weak, in a manner to be defined, and the levitated sphere is small compared to the capillary length of the liquid bath; we will also assume that the densities of the bath and sphere are comparable, though not necessarily the same. Crucial to the formation of  capillary waves in the vapour film, the small-sphere restriction  does not imply that gravity effects are negligible there, unlike in the small-drop limit of the classical Leidenfrost scenario involving a flat substrate \cite{Pomeau:12}.

Our assumption that the levitating object is a non-deformable sphere, rather than a deformable liquid drop, is made for the sake of simplicity of the analysis. In the inverse-Leidenfrost scenario, where the liquid bath evaporates, this assumption is adequate when the levitating object is either an inert solid particle (e.g., metallic particles were used in \cite{Hall:69} and polyethylene particles were used in \cite{Gauthier:19}) or a frozen, high-surface-tension or surface-contaminated liquid drop \cite{Adda:16}. In the case of a heated liquid bath, the levitated spherical particle may represent a high-surface-tension evaporating drop, or a subliming mass of dry ice (the solid form of carbon dioxide) as used to demonstrate Leidenfrost levitation of solid masses above solid substrates \cite{Lagubeau:11}. 

Asymptotic analyses and numerical models of Leidenfrost drops \cite{Pomeau:12,Sobac:14,Maquet:16,Adda:16,Van:19} and closely related configurations \cite{Duchemin:05,Lister:08,Snoeijer:09} have typically assumed a lubrication-type description of the vapour film as an approximate starting point. We adopt this approach here. In the present case, however, no need arises  for the usual `patching' of the lubrication approximation of the vapour film with a separate hydrostatic approximation of the region exterior to the vapour film. Rather, with the spherical geometry of the levitating object known \textit{a priori}, the rapid radial attenuation of the vapour flow outside the film can be exploited towards formulating a `unified' model that is applicable everywhere, in a leading-order sense.

In deriving this unified lubrication model we shall in particular assume that the slope of the height profile of the liquid bath is everywhere small.
Similar small-slope assumptions have been used to model drops levitated above a flat solid substrate \cite{Snoeijer:09,Pomeau:12,Sobac:14} but not to model large drops levitated above a heated liquid bath \cite{Van:19} and other levitation scenarios \cite{Duchemin:05,Lister:08} where the film and substrate are significantly curved. (The lubrication model in \cite{Maquet:16} of drops levitated above a heated liquid bath implicitly assumes small slopes despite that assumption being generally inadequate for the moderately sized drops considered therein.) Here, we shall verify \textit{a posteriori} that the asymptotic solutions obtained for small levitated spheres and weak evaporation are consistent in this regard. 

The `unified small-slope model' we shall employ as a starting point for our asymptotic analysis involves only two independent dimensionless numbers, a modified Bond number $\mathcal{B}$ and a modified evaporation number $\mathcal{E}$. (The modifications alluded to allow us to suppress explicit dependence upon the liquid-solid density ratio.) In accordance with the goal of this study, our  analysis will focus on the small-sphere regime $\mathcal{B}\ll1$. The evolution of the film morphology with increasing sphere radius will be seen to correspond to formally infinite sequences of distinguished limits associated with the  smallness of $\mathcal{E}$ relative to $\mathcal{B}$.

\section{Problem formulation}\label{sec:formulation}
\subsection{Physical model}\label{ssec:pmodel}
\begin{figure}[t!]
\begin{center}
\includegraphics[scale=0.173]{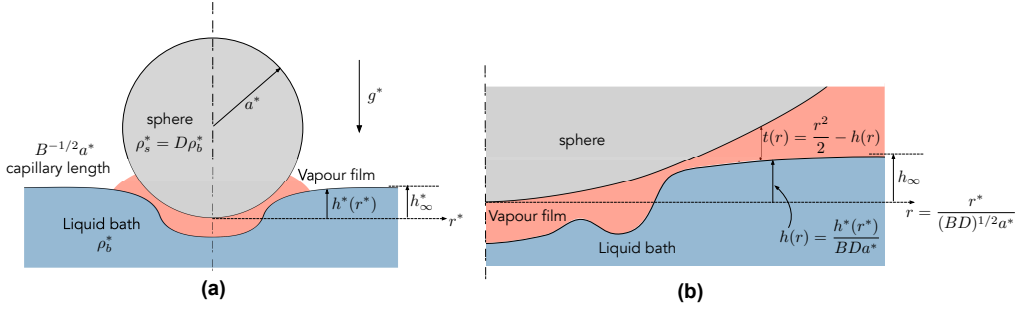}
\caption{(a) Dimensional schematic of the physical problem formulated in \S\ref{ssec:pmodel}. (b) Dimensionless schematic corresponding to the `unified small-slope' model developed in \S\ref{ssec:nondim}.}
\label{fig:schematic}
\end{center}
\end{figure}

We consider the levitation of a stationary rigid sphere  (radius $a^*$, density $\rho_s^*$) above an infinite liquid bath (density ${\rho}_b^*$). The levitation is sustained by a layer formed of the vapour of either the liquid bath or the sphere (thermal conductivity $k^*$, density $\rho^*$, viscosity $\mu^*$). Assuming axial symmetry, we denote by $h^*(r^*)$ the height of the liquid bath measured relative to the bottom of the sphere, with $r^*$ the radial distance from the symmetry axis; we denote by $h^*_{\infty}$ the unperturbed height of the liquid bath and refer to it as the sphere displacement (Fig.~\ref{fig:schematic}a). The liquid-vapour interfacial tension is $\gamma^*$ and the gravitational acceleration is $g^*$. We use a superscript asterisk to indicate a dimensional quantity. 

Vapour production in this system is associated with a temperature difference (absolute magnitude $\Delta T^*$) between the sphere and bath, both of which assumed isothermal on account of their thermal conductivities being sufficiently high. Specifically, we assume that phase change occurs at the boundary of the cooler object, which is at its temperature for either evaporation or sublimation. This translates into the following condition on the vapour velocity at the phase-changing boundary \cite{Pomeau:12}:
\begin{equation}\label{evap velocity}
\bv^* = \frac{k^*}{\rho^*l^*}(\bn\bcdot\bnabla^* T^*)\bn,
\end{equation}
where $l^*$ is the pertinent latent heat, $\bn$ is a normal unit vector pointing into the vapour layer and $\bnabla^* T^*$ is the local vapour-temperature gradient. We assume that heat is transported solely by conduction through the vapour layer. We additionally assume that all of the physical quantities mentioned above can be considered constant on the time scale of interest and that the dynamical effect of the liquid-bath flow can be neglected  on the assumption that the liquid is much more viscous than the vapour.

The problem thus consists of (i) the Stokes equations governing the flow in the vapour layer (supposed to fill the entire fluid domain), subject to \eqref{evap velocity} on the phase-changing surface and zero velocity on the other; (ii) the dynamical boundary condition at the liquid-vapour interface, in which the dynamical effect of the liquid bath enters through capillarity and the hydrostatic pressure
\begin{equation}
-\rho_b^*g^*\left(h^*(r^*)-h^*_{\infty}\right),
\end{equation}
which is taken to vanish at the unperturbed bath height; 
(iii) the condition that the gravity force on the sphere is balanced by the vertical hydrodynamic force exerted by the vapour layer on the sphere; and (iv) the steady heat-conduction problem in the vapour layer which determines the distribution of the temperature gradient in \eqref{evap velocity}. 

The above problem involves three independent dimensionless groups:
\refstepcounter{equation}
$$
\label{BED}
B=\frac{{\rho}_b^*g^* {a^*}^2}{\gamma^*}, \quad E=\frac{\mu^*\lambda^*}{\gamma^*a^*}, \quad 
D=\frac{{\rho_s^*}}{{\rho}_b^*},\eqno{(\theequation{\mathrm{a}\!\!-\!\!\mathrm{c}})}
$$
where $\lambda^*=k^*\Delta T^* / \rho^*l^*$. Here $B={a^*}^2/{\ell_c^*}^2$ is the Bond number, namely the ratio squared of the sphere radius to the liquid-bath capillary length $\ell_c^*=(\gamma^*/\rho_b^*g^*)^{1/2}$, $E$ is an evaporation number  and $D$ is a density ratio. (It will be convenient to refer to $E$ as an evaporation number regardless of whether vapour is being produced by evaporation or sublimation.) In the inverse-Leidenfrost scenario where an inert sphere levitates above an evaporating liquid-nitrogen bath, typical values are $\ell_c^* \approx 1$ mm and $a^* E \approx 10$ nm \cite{Adda:16}. 

\subsection{Unified small-slope model} \label{ssec:uni}
Our interest is in the small-sphere and weak-evaporation regime where $B$ and $E$ are both small, while $D$ is of order unity. Below, we intuitively formulate a `unified small-slope' lubrication model to describe this regime. This model will serve as a starting point for a systematic asymptotic analysis in subsequent sections. 

We assume that the slope of the bath height $dh^*/dr^*$ is everywhere small. The dynamical condition at the liquid-vapour interface can then be approximated as
\begin{equation}\label{dyn star}
p^*=-\rho_b^*g^*(h^*-{h_{\infty}}^*)+\frac{{\gamma^*}}{r^*}\frac{d}{dr^*}\left(r^*\frac{dh^*}{dr^*}\right),
\end{equation}
where $p^*$ is the vapour pressure field evaluated at the liquid-vapour interface. Considering \eqref{dyn star} as an equation for the bath height $h^*$, we have the boundary condition 
\begin{equation}\label{dh zero star}
\frac{dh^*}{dr^*}=0 \quad \text{at} \quad r^*=0,
\end{equation}
which follows from axial symmetry, and the far-field condition
\begin{equation}\label{h far star}
h^*\to h_{\infty}^* \quad \text{as} \quad r^*\to\infty.
\end{equation}

Since \eqref{dyn star} involves $p^*$, we need to add equations governing the vapour flow in order to close the problem for $h^*$. In particular, we anticipate a thin-film region whose radial extent, say of order $\ell_f^*$, is  large relative to the thickness of the film $t^*$ yet small relative to $a^*$. In that region, the sphere boundary can be approximated by a paraboloid such that
\begin{equation}\label{t def star}
t^*=\frac{1}{2a^*}{r^*}^2 - h^*,
\end{equation}
where $dt^*/dr^*\ll1$ follows as a consequence of \eqref{t def star} and the above assumptions.

Given that both $h^*$ and $t^*$ are slowly varying, we can employ a standard lubrication approximation to describe the vapour flow in the thin-film region. Consistent with this approximation, we use $\bn\bcdot \bnabla T^*\approx \Delta T^*/t^* $ in the boundary condition \eqref{evap velocity} since the vapour temperature varies approximately linearly across the thin film (for a detailed derivation, see \cite{Pomeau:12}). We thereby find the Reynolds, or flux-conservation, equation 
\begin{equation}\label{Reynolds star}
\frac{1}{r^*}\frac{d}{dr^*}\left(r^*q^*\right)=\frac{\lambda^*}{t^*}
\end{equation}
where the flux density $q^*$ is defined as
\begin{equation}\label{flux star}
q^*=-\frac{{t^*}^3}{12\mu}\frac{dp^*}{dr^*},
\end{equation}
and the source term on the right-hand-side of \eqref{Reynolds star} accounts for vapour production at the phase-changing boundary. 
(Note that the net volumetric flux out of a cylindrical control volume of radius $r^*$, which is collinear with the symmetry axis, is given by $2 \pi r^* q^*$.) 
From axial symmetry, we have the boundary condition
\begin{equation}\label{q zero star}
q^*=0 \quad \text{at} \quad r^*=0.
\end{equation}

Recall that \eqref{t def star}--\eqref{q zero star} were meant to describe the thin-film region. We anticipate, however, that $p^*$ attenuates fast enough for $r\gg \ell_f^*$ such that the vapour pressure term in the dynamical condition \eqref{dyn star} can be safely neglected in that region. 
This suggests extending the lubrication equations \eqref{t def star}--\eqref{q zero star} to all $r^*$, to be solved in conjunction with the far-field condition
\begin{equation}\label{p far star}
p^*\to0 \quad \text{as} \quad r^*\to\infty
\end{equation}
and the approximate vertical-force balance
\begin{equation}\label{force star}
\int_0^{\infty} p^*r^*\,dr^*=\frac{2}{3}\rho_b^* g^* {a^*}^3.
\end{equation}
In the latter balance, the integral 
represents the vertical pressure force contributed by the thin-film region, assuming that $p^*$ attenuates sufficiently fast that the upper range of integration can be taken to infinity; viscous stresses as well as the slope of the sphere boundary have been neglected 
consistently with previous assumptions.

We will refer to \eqref{dyn star}--\eqref{force star} as the unified small-slope model. We use the term `unified' rather than `uniform' as we do not expect the solutions to constitute uniform asymptotic approximations in any given limit. Rather, the model is meant to provide solutions that constitute, for any fixed $r^*$, leading-order approximations for $t^*$ in the small-sphere and weak-evaporation regime. 
It is a simple matter to verify \textit{a posteriori} that the asymptotic solutions we shall derive based on the unified small-slope model are indeed consistent with the assumptions on which that model is predicated. 

\subsection{Dimensionless equations}\label{ssec:nondim}
Dimensional analysis of the unified small-slope model furnishes just two independent dimensionless groups, which can be chosen as the modified Bond and evaporation numbers
\refstepcounter{equation}
$$
\label{BE modified}
\mathcal{B}=D^{1/2}B, \quad \mathcal{E}=\frac{E}{(BD)^3}, 
\eqno{(\theequation{\mathrm{a},\!\mathrm{b}})}
$$
respectively.
Note that $\mathcal{B}\propto {a^*}^2$ and $\mathcal{E}\propto {1/{a^*}^7}$, when varying the sphere radius $a^*$ with all other physical parameters fixed. 
In what follows, we adopt a dimensionless convention where the radial coordinate $r^*$ is normalised by $(BD)^{1/2}a^*$, the height of the liquid bath $h^*$, the sphere displacement $h^*_{\infty}$ and the thickness $t^*$ by $BDa^*$,  the vapour pressure $p^*$ by $\gamma^*/a^*$ and the flux density $q^*$ by $(BD)^{5/2}\gamma^{*} a^{*}/\mu^*$; the associated normalised quantities are denoted similarly but with the asterisk omitted. In particular, the relation \eqref{t def star} between the thickness and the bath height becomes
\begin{equation}\label{t def}
t=\frac{1}{2}r^2-h.
\end{equation}
A dimensionless schematic is shown in Fig.~\ref{fig:schematic}b. We note that the modified evaporation number $\mathcal{E}$ differs from those used in preceding analyses of the Leidenfrost effect \cite{Sobac:14,Maquet:16,Van:19}. 

In dimensionless form, the unified small-slope model consists of the lubrication equations 
\refstepcounter{equation}
$$
\label{q eqs}
\frac{1}{r}\frac{d}{dr}\left(rq\right)=\frac{\mathcal{E}}{t}, \quad q=-\frac{{t}^3}{12}\frac{dp}{dr};
\eqno{(\theequation{\mathrm{a},\!\mathrm{b}})}
$$
the dynamic condition
\begin{equation}\label{p h eq}
p=-\mathcal{B}^2\left(h-h_{\infty}\right)+\frac{1}{r}\frac{d}{dr}\left(r\frac{dh}{dr}\right),
\end{equation}
which in terms of thickness can be written 
\begin{equation}\label{p t eq}
p=2-\frac{\mathcal{B}^2}{2}r^2+\mathcal{B}^2\left(t+h_{\infty}\right)-\frac{1}{r}\frac{d}{dr}\left(r\frac{dt}{dr}\right);
\end{equation}
the boundary conditions
\refstepcounter{equation}
$$
\label{bcs at zero}
\frac{dh}{dr}=0 \,\, \left(\text{or }\frac{dt}{dr}=0\right), \quad q=0 \quad \text{at} \quad r=0;
\eqno{(\theequation{\mathrm{a},\!\mathrm{b}})}
$$
the far-field conditions
\refstepcounter{equation}
$$
\label{bcs at infinity}
p\to0, \quad h\to h_{\infty} \quad \text{as} \quad r\to\infty;
\eqno{(\theequation{\mathrm{a},\!\mathrm{b}})}
$$
and the vertical force balance 
\begin{equation}\label{force}
\int_0^{\infty}pr\,dr=\frac{2}{3}.
\end{equation}

In the remainder of the paper, we asymptotically analyse the unified small-slope model
in the small-sphere regime $\mathcal{B}\ll1$. We shall see that this regime is comprised of formally infinite sequences of distinguished limits, the limits being defined by the smallness of $\mathcal{E}$ relative to $\mathcal{B}$. We will initially assume that the modified evaporation number $\mathcal{E}$ is order unity and find that the unified small-slope model breaks down for large enough $\mathcal{E}$. Subsequently, we shall consider limits corresponding to relatively smaller and smaller $\mathcal{E}$.

We note that the unified small-slope model is fairly straightforward to solve numerically, for example using Matlab routines for boundary-value problems \cite{MATLAB:2020}. (In order to accurately impose the far-field conditions, it is useful to develop local asymptotics of the fields for large $r$.) Numerical solutions obtained in this manner will be used throughout the paper for the sake of comparison with asymptotic  approximations. At this stage, the reader may wish to refer to Figs.~\ref{fig:firstprofiles}(a), \ref{fig:firstprofiles}(b), \ref{fig:2lim} and \ref{llim} showing film profiles calculated for $\mathcal{B}=0.1$ and $\mathcal{E}=0.1,10^{-6},10^{-22}$ and $10^{-29}$, respectively. With respect to the morphological evolution discussed in the introduction, Fig.~\ref{fig:firstprofiles} demonstrates the initial stage of the evolution where a neck-bubble structure forms, whereas Fig.~\ref{llim} demonstrates the final stage of the evolution where the film consists mostly of a uniform-thickness region which expands radially. Our analysis will theoretically illuminate the linkage between these stages.

While values of $\mathcal{E}$ smaller than around $10^{-6}$ are unrealistic (see estimates below \eqref{BED}), the need to consider exceedingly small $\mathcal{E}$ in order to observe the late stages of the evolution is a consequence of fixing $\mathcal{B}$ in the above numerical examples. We do this so that the formally small parameters in our analysis, $\mathcal{B}$ and $\mathcal{E}$, remain numerically small throughout the evolution.  
Alternatively, a similar evolution can be observed by increasing the sphere radius $a^*$ with all other parameters fixed, in which case $\mathcal{B}$ increases while $\mathcal{E}$ decreases (as shown in \cite{Maquet:16} for drops levitated above a liquid substrate). While in that scenario $\mathcal{E}$ and $\mathcal{B}$ maintain  physically reasonable values, $\mathcal{B}$ is not so small during the later stages of the evolution. Thus, we expect only a qualitative correspondence between our asymptotic analysis and the later stages of the evolution as they would be observed numerically or experimentally for typical values of the physical parameters. Moreover, in that scenario our unified small-slope model is unlikely to be accurate since, as will become clear, the assumptions underlying it imply $\mathcal{B}\ll1$. 

\section{Thin-film and formation of neck-bubble morphology}\label{sec:formation}
\subsection{Thin-film region}\label{ssec:thinfilm}
In this section we consider the unified small-slope model in the limit $\mathcal{B}\to0$, with $\mathcal{E}\simeq 1$. (Henceforth, we use $\simeq$ to denote asymptotic order.) We shall later refer to this regime as the first distinguished limit.

We start by posing the straightforward expansions   
\refstepcounter{equation}
$$
\label{First expansions}
 h=h_0(r)+\cdots, \quad t=t_0(r)+\cdots, \quad q= q_0(r) + \cdots, \quad p= p_0(r) + \cdots,
\eqno{({\theequation}\mathrm{a}\!\!-\!\!\mathrm{d})}
$$
where, from \eqref{t def},
\begin{equation}\label{First t def}
t_0=\frac{1}{2}r^2-h_0.
\end{equation} 

We now list the equations satisfied by the leading-order fields. From \eqref{q eqs} and \eqref{p h eq}, we have
\refstepcounter{equation}
$$
\label{First eqs}
\frac{1}{r}\frac{d}{dr}\left(rq_0\right)=\frac{\mathcal{E}}{t_0}, \quad 
q_0=-\frac{t_0^3}{12}\frac{dp_0}{dr}, \quad
p_0=\frac{1}{r}\frac{d}{dr}\left(r\frac{dh_0}{dr}\right),
\eqno{(\theequation \mathrm{a}\!\!-\!\!\mathrm{c})}
$$
which are identical to the equations of the unified small-slope model except for the absence of  gravity terms in the approximate dynamic condition (\ref{First eqs}c). 
From \eqref{bcs at zero}, we have the boundary conditions
\refstepcounter{equation}
$$
\label{First bcs}
\frac{dh_0}{dr}=0, \quad q_0=0 \quad \text{at} \quad r=0,
\eqno{(\theequation \mathrm{a},\!\mathrm{b})}
$$
while (\ref{bcs at infinity}a) gives the attenuation condition 
\begin{equation}\label{First p far}
p_0\to0 \quad \text{as} \quad r\to\infty. 
\end{equation}

We have yet to refer to the far-field condition (\ref{bcs at infinity}b) on the bath height and to the force constraint \eqref{force}. With respect to the former, it is noted that the attenuation of $p_0$ together with (\ref{First eqs}c) implies the far-field behaviour 
\begin{equation}\label{First h far}
h_0 \sim A\ln r \quad \text{as} \quad r\to\infty,
\end{equation}
where $A$ is a constant. With $h_0$ growing logarithmically, the far-field condition (\ref{bcs at infinity}b) cannot be satisfied. The issue is that at large distances, comparable to the capillary length, or $r\simeq 1/\mathcal{B}$ in our normalisation scheme, the gravity terms neglected in \eqref{p h eq} become important. Expansions \eqref{First expansions} are accordingly interpreted as `outer' approximations, i.e., for $r$ fixed, which describe the thin-film region. In principle, the value of $A$ in \eqref{First h far} comes from matching with an `inner' hydrostatic region. It is clear, however, that the pressure integral in the force balance  \eqref{force} is dominated by the thin-film region, whereby  \eqref{force} yields 
\begin{equation}\label{First force}
\int_0^\infty p_0 r\,dr=\frac{2}{3}.
\end{equation}
Hence $A$ can be readily calculated by substituting (\ref{First eqs}c) and \eqref{First h far} into \eqref{First force}. This gives \begin{equation}\label{First hfar}
A=\frac{2}{3}.
\end{equation}

The leading-order thin-film problem is now closed; it is, in fact, mathematically identical to that formulated in \cite{Pomeau:12} for a small Leidenfrost drop levitated above a flat solid substrate. In the present problem, however, it remains to satisfy the far-field condition (\ref{bcs at infinity}b) on the bath height; this does not imply an overdetermined problem as that condition involves the yet unknown sphere displacement $h_{\infty}$. In the next subsection, we calculate $h_{\infty}$ by matching the thin-film region with the hydrostatic region. 
\begin{figure}[b!]
\centering
\includegraphics[trim={2.75cm 2cm 0 1cm},scale=0.345]{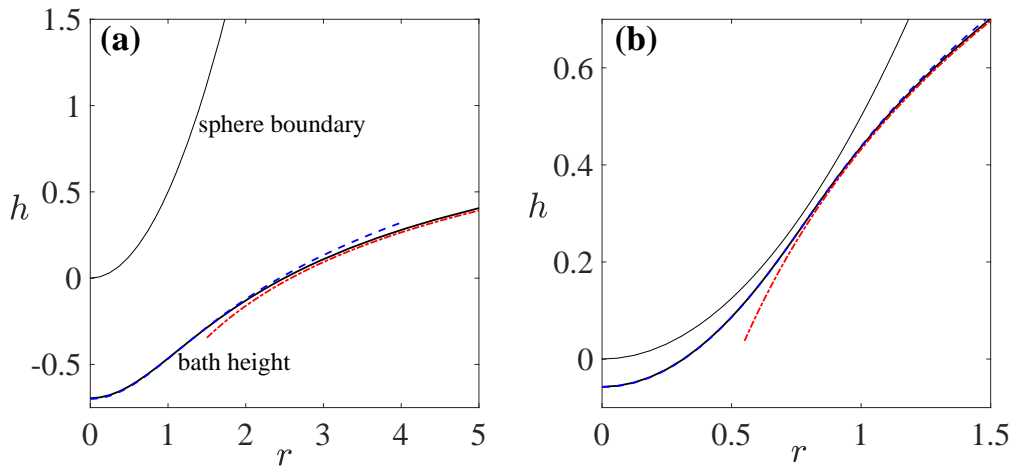}
\caption{The height profile $h(r)$ for $\mathcal{B}=0.1$ and (a) $\mathcal{E}=0.1$; (b) $\mathcal{E}=10^{-6}$. The numerical solution of the unified small-slope model (lower solid line) is compared with asymptotic approximations derived in \S\ref{sec:formation} in the first distinguished limit for the thin film region (dashed line) and hydrostatic region (dash-dotted line). The sphere, locally approximated by a paraboloid, is represented by the upper solid line. As $\mathcal{E}$ decreases, the sphere levitates closer to the bath and the thin film decomposes into a bubble region bounded by a narrow and thin neck region (see \S\ref{ssec:neckbubble}).}
\label{fig:firstprofiles}
\end{figure}

\begin{figure}[t!]
\centering
\includegraphics[trim={4cm 0cm 3cm 0},scale=0.28]{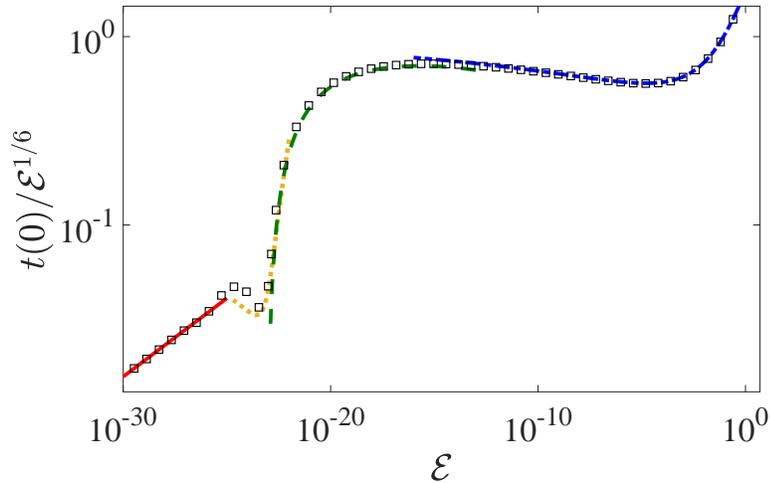}
\caption{Scaled film thickness at the origin as a function of $\mathcal{E}$, for $\mathcal{B}=0.1$. Square symbols: numerical solution of the unified small-slope model (\S\ref{ssec:nondim}). Dash-dotted curve: approximation in the first distinguished limit (\S\ref{ssec:thinfilm}). Dashed curve: approximation in the second distinguished limit (\S\ref{ssec:second}). Dotted curve: approximation in the third distinguished limit (\S\ref{ssec:third}). Solid curve: approximation valid throughout the second formally infinite sequence of distinguished limits (\S\ref{Sec:uniform1}).  }
\label{fig:thick}
\end{figure}

It is straightforward to solve the leading-order thin-film problem  numerically. In particular, unlike the unified small-slope model, the present reduced problem does not involve multiple length scales, at least for values of $\mathcal{E}$ on the order of unity (see \S\ref{ssec:neckbubble}). A comparison between the two schemes is presented in Fig.~\ref{fig:firstprofiles}, where the bath-height profile is plotted for $\mathcal{B}=0.1$ and two values of $\mathcal{E}$, $0.1$ and $10^{-6}$. Further comparison is presented in Fig.~\ref{fig:thick}, which shows the variation with $\mathcal{E}$ of the film thickness at the origin. 

We will see that the specific output of the thin-film problem that is required for matching with the  hydrostatic region is the limit 
\begin{equation}\label{beta def}
\beta = \lim_{r\to\infty}\left(h_0-\frac{2}{3}\ln r\right),
\end{equation}
which corresponds to the constant order-unity correction to the far-field expansion \eqref{First h far}. 
Like the thin-film problem itself, $\beta$ is a function of $\mathcal{E}$ alone. Analytical approximations for $\beta$ can be deduced for small and large $\mathcal{E}$. In the latter limit, corresponding to strong evaporation, or relatively small spheres, the bath height is approximately constant and equal to $\beta$. The geometry is then simply that of a sphere above a vertically displaced flat bath. In that case, the approximation 
\begin{equation}\label{beta E large}
\beta \sim -\frac{3}{2}\mathcal{E}^{1/2} \quad \text{as} \quad \mathcal{E}\to\infty
\end{equation}
follows from the analysis in \cite{Pomeau:12} of an analogous limit (there appears to be a typo in the approximation obtained therein).
 The diametric limit $\mathcal{E}\to0$, which corresponds to weak evaporation, or relatively large spheres, will be discussed in \S\ref{ssec:neckbubble} and in a wider context for the remainder of this paper. For the moment, we simply state the result 
\begin{equation}\label{beta E small}
\beta \to \frac{1}{3}\left(\ln\frac{3}{2}+1\right) \quad \text{as} \quad \mathcal{E}\to0,
\end{equation}
which will fall out from the analysis in \S\ref{sec:onset}. (As explained in \cite{Sobac:14}, the weak-evaporation analysis in \cite{Pomeau:12} is erroneous. Thus, \eqref{beta E small} cannot be extracted from that analysis.) The function $\beta(\mathcal{E})$, as obtained from numerically solving the thin-film problem, is shown in Fig.~\ref{fig:beta} along with the analytical approximations \eqref{beta E large} and \eqref{beta E small}.

\subsection{Hydrostatic region, matching and sphere displacement}\label{ssec:hydro}
To study the hydrostatic region we define the strained coordinate 
\begin{equation}
\hat{r}=\mathcal{B}r.
\end{equation}
Given \eqref{First h far}, it is clear that $h\simeq 1$ in the hydrostatic region so that the gravity and capillarity terms in the dynamic condition \eqref{p h eq} are $\simeq \mathcal{B}^2$. In contrast, since $p_0\simeq 1/r^6$ as $r\to\infty$, as readily follows from local analysis of the thin-film problem, the pressure term in \eqref{p h eq} is $\simeq \mathcal{B}^6$. (These scalings are accurate  to within  logarithmic factors.) Thus, in the hydrostatic region the dynamic condition provides an equation for the bath height which is uncoupled from the vapour flow, except through asymptotic matching. 

In light of the above, we pose the expansions
\refstepcounter{equation}
$$
\label{hydro h exp}
h = \hat{h}(\hat{r}) + \cdots, \quad h_{\infty} = \hat{h}_{\infty} + \cdots.
\eqno{({\theequation}\mathrm{a},\!\mathrm{b})}
$$
The leading-order field $\hat{h}$ satisfies 
\begin{equation}\label{phi eq}
\frac{1}{\hat{r}}\frac{d}{d\hat{r}}\left(\hat{r}\frac{d\hat{h}}{d\hat{r}}\right)-\hat{h} = -\hat{h}_{\infty}, 
\end{equation}
which follows from the dynamic condition \eqref{p h eq}, the far-field condition
\begin{equation}\label{phi far}
\hat{h}\to \hat{h}_{\infty}  \quad \text{as} \quad \hat{r}\to\infty,
\end{equation}
which follows from (\ref{bcs at infinity}b), and matching conditions in the limit $\hat{r}\to0$. 

The solution to \eqref{phi eq} and \eqref{phi far} consists of a superposition of the constant particular solution $\hat{h}_{\infty}$ and a homogeneous solution that decays as $\hat{r}\to\infty$. The latter solution is proportional to the modified Bessel function of the second kind $K_0(\hat{r})$ that possesses the asymptotic behaviour 
\begin{equation}
K_0(\hat{r})\sim -\ln \hat{r} + \ln 2 - \gamma_E \quad \text{as} \quad \hat{r}\searrow 0,
\end{equation}
where $\gamma_E\doteq 0.5772\ldots$ is the Euler--Gamma constant \cite{Abramowitz:book}. Using this behaviour, straightforward matching between the thin-film and hydrostatic regions implies the solution
\begin{equation}\label{hydro h}
\hat{h} = \hat{h}_{\infty} -\frac{2}{3} K_0(\hat{r}), 
\end{equation}
with the leading-order sphere displacement $\hat{h}_{\infty}$ provided as 
\begin{equation}\label{varphi}
\hat{h}_{\infty}=\frac{2}{3}\ln\frac{2}{\mathcal{B}}-\frac{2}{3}\gamma_E + \beta ,
\end{equation}
wherein $\beta$ is the function of $\mathcal{E}$ calculated from the thin-film problem (cf.~\eqref{beta def}).

\begin{figure}[t!]
\includegraphics[trim={0 2cm 0 0},scale=0.33]{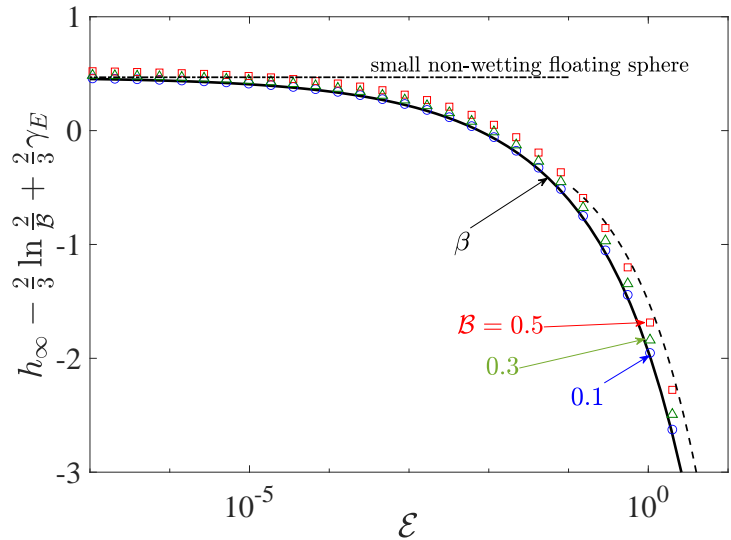}
\caption{Sphere displacement $h_{\infty}$ as a function of $\mathcal{E}$, after subtracting the $\mathcal{B}$ dependence anticipated from the  asymptotic approximation \eqref{varphi}. Numerical solutions of the unified small-slope model for several values of $\mathcal{B}$ (symbols) collapse on the asymptotic prediction based on \eqref{varphi}, which following the  subtraction equals the function $\beta(\mathcal{E})$ defined by the thin-film problem of \S\ref{ssec:thinfilm}. The large-$\mathcal{E}$ approximation \eqref{beta E large} and small-$\mathcal{E}$ approximation \eqref{beta E small} of that function are depicted by the dashed and dash-dotted curves, respectively. The latter approximation also describes a small non-wetting sphere floating on a liquid bath.}
\label{fig:beta}\end{figure}

In Fig.~\ref{fig:beta}, we validate the asymptotic prediction \eqref{varphi} against numerical solutions of the unified small-slope model. The numerical values for the sphere displacement, for several small values of $\mathcal{B}$, are seen to collapse on the asymptotic prediction once the $\mathcal{B}$ dependence anticipated from \eqref{varphi} is subtracted. Analytical approximations for the sphere displacement in the limits of large and small $\mathcal{E}$ can be obtained by substituting the corresponding approximations \eqref{beta E large} and \eqref{beta E small} for $\beta$ into \eqref{varphi}. In particular, substituting the latter gives the limit
\begin{equation}\label{floating}
\hat{h}_{\infty}\to \frac{2}{3}\ln\frac{
{\sqrt{6}}}{\mathcal{B}}+\frac{1}{3}-\frac{2}{3}\gamma_E \quad \text{as} \quad \mathcal{E}\to0,
\end{equation}
which corresponds to the displacement of a small non-wetting sphere floating on a liquid bath \cite{Cooray:17,Galeano:20}.

We conclude this subsection with the following remarks:
\begin{enumerate}
\item Note the appearance in \eqref{varphi} of the logarithm of the small parameter $\mathcal{B}$. Except where stated otherwise, we shall collect together terms that are of the same algebraic order in $\mathcal{B}$. When technically possible, this convention allows obtaining asymptotic approximations that are algebraically, rather than logarithmically, accurate, without the need to carry out the analysis at successive logarithmic orders \cite{Hinch:book}. Consistently with this convention, the symbol $\simeq$ should be interpreted as asymptotic order ignoring logarithmic factors.
\item At this stage, it may not be clear why we introduced a unified small-slope model, only to asymptotically decompose that model at the first opportunity into thin-film and hydrostatic regions. We could have also arrived at this picture by a direct matched asymptotics analysis of the original physical problem formulated in \S\ref{ssec:pmodel}. Nonetheless, the unified formulation will prove useful in the remainder of the paper, where we will study a large number of distinguished limits associated with the relative smallness of $\mathcal{B}$ and $\mathcal{E}$, the analysis in this section corresponding to just the first of these limits. Analysing each of these limits directly from the original formulation is technically possible but very tedious and repetitive. 
\item As discussed in \S\ref{ssec:firstBD}, the reduced model derived in the present section loses  validity at sufficiently small $\mathcal{E}$. Nonetheless, the result \eqref{varphi} for the sphere displacement, including its limiting value \eqref{floating}, remains valid. This is confirmed by the analysis in the subsequent sections and is also evident from the comparison presented in  Fig.~\ref{fig:beta}. 
\item From \eqref{beta E large} it can be deduced that $h_0\simeq\mathcal{E}^{1/2}$ for large $\mathcal{E}$. It then follows from \eqref{First t def} that $dt_0/dr\simeq\mathcal{E}^{1/4}$ in that limit. Using the scaling factors in \S\ref{ssec:nondim}, the corresponding scaling of $dt^*/dr^*$ is $(BD)^{1/2} \mathcal{E}^{1/4}$. For $D$ of order unity, the latter scaling implies that the unified small-slope model breaks down for $ \mathcal{E}\simeq 1/\mathcal{B}^{2}$.
\end{enumerate}

\subsection{Formation of neck-bubble morphology}\label{ssec:neckbubble}
Consider the thin-film problem in the limit $\mathcal{E}\to0 $. As seen in Fig.~\ref{fig:firstprofiles}, the vapour film decomposes into a very thin central bubble region bounded by a narrow annular neck region which is even thinner. This morphology is analogous to that familiar in the classical Leidenfrost effect \cite{Sobac:14}. We note, in particular, that for small $\mathcal{E}$ the origin becomes a maximum of the thickness profile. 

Let $L_B$ and $T_B$ be the orders of the radial extent and thickness of the vapour bubble, respectively. Assuming that $T_B\ll L_B^2$, namely that the bubble is thin compared to the vertical variation of the sphere boundary over that region, the pressure in the bubble region must approximately equal the capillary pressure associated with the sphere's curvature, i.e.,
\begin{equation}\label{bubble pressure}
p_0  \sim 2.
\end{equation}
Assuming that the bubble region dominates the pressure integral in the force balance \eqref{First force}, we have that $L_B=1$; specifically, the vapour bubble must terminate  at $r=l$, where 
\begin{equation}\label{bubble extent}
l= \sqrt{2/3}.
\end{equation}
Let $Q_B$, $P'$ and $\Delta P$ be the orders of the flux density, pressure correction and pressure variation, respectively, in the bubble region. With reference to the flux-conservation equation (\ref{First eqs}a), it is clear that the vapour-production term on the right-hand side of that equation cannot be negligible in the bubble region; otherwise, the flux density would be singular at the origin. Thus, the thin-film equations \eqref{First eqs} imply
\refstepcounter{equation}
$$\label{First BD bubble scalings}
Q_B=\frac{\mathcal{E}}{T_B}, \quad P'=T_B, \quad \Delta P 
=\frac{\mathcal{E}}{T_B^5},
\eqno{(\theequation \mathrm{a}\!\!-\!\!\mathrm{c})}
$$
where the scaling $T_B\ll1$ remains to be determined.

Consider now the neck region at the edge of the bubble region. Denote the orders of the neck's radial width, thickness and flux density by $L_N$, $T_N$ and $Q_N$, respectively. Since $p_0$ is clearly order unity and not constant in the neck region, (\ref{First eqs}b) and (\ref{First eqs}c) imply 
\refstepcounter{equation}
$$
\label{First BD neck scalings pre}
T_N=L_N^2, \quad Q_N=L_N^5.
\eqno{(\theequation \mathrm{a},\!\mathrm{b})}
$$
It is also evident that $L_N\ll1$ and $T_N\ll T_B$. 

We now relate the bubble and neck scalings based on matching considerations. First, using $L_N\ll1$, the dynamical condition (\ref{First eqs}c) simplifies in the neck region as
\begin{equation} \label{neck p approx}
\frac{d^2t_0}{dr^2}\approx  2-p_0.
\end{equation}
Since $p_0\to 2$ towards the bubble region (cf.~\eqref{bubble pressure}), $t_0$ grows linearly in that direction. Matching $dt_0/dr$ thus gives $T_B/L_B=T_N/L_N$, i.e., $T_B=L_N$. Second, we argue that $Q_B=Q_N$. Without vapour production, this would be obvious from continuity. With vapour production, inspection of the flux-conservation equation (\ref{First eqs}a) together with the fact that the bubble thickness vanishes linearly towards the neck means that $Q_B$ and $Q_N$ can differ only by a factor of logarithmic order in $\mathcal{E}$. Thus, up to such logarithmic factors, we find
\refstepcounter{equation}
$$
\label{First BD neck scalings}
L_N = \mathcal{E}^{1/6}, \quad T_N = \mathcal{E}^{1/3}, \quad Q_N = \mathcal{E}^{5/6},
\eqno{(\theequation \mathrm{a}\!\!-\!\!\mathrm{c})}
$$
and 
\refstepcounter{equation}
$$
\label{First BD bubble scalings}
T_B=\mathcal{E}^{1/6}, \quad Q_B=\mathcal{E}^{5/6}, \quad P'_B=\mathcal{E}^{1/6}, \quad \Delta P_B = \mathcal{E}^{1/3}.
\eqno{(\theequation \mathrm{a}\!\!-\!\!\mathrm{d})}
$$
It follows from the above scalings that the pressure correction $p_0-2$ in the bubble region is approximately uniform.
Furthermore, vapour production in the neck is not negligible.  

In principle, the decomposition of the thin-film problem at small $\mathcal{E}$ includes a third region, exterior to the neck region. We shall refer to it as the transition region as it bridges the neck region and the far-field behaviour \eqref{First h far}. We denote the orders of the radial extent, thickness, flux density and pressure in the transition region by $L_T,T_T,Q_T$ and $P_T$, respectively. Anticipating that $p_0$ decreases when crossing the neck in the outward radial direction, the approximate dynamical condition \eqref{neck p approx} implies that $t_0$ grows quadratically in that direction away from the neck. Order-of-magnitude matching of $d^2t_0/dr^2$ accordingly gives $T_T/L_T^2=T_N/L_N^2$, or $T_T=L_T^2$. Together with the height-thickness relation \eqref{First t def} and far-field condition \eqref{First h far}, this implies $L_T=T_T=1$. Similarly, matching the flux density gives $Q_T=\mathcal{E}^{5/6}$, whereby the flux-pressure relation (\ref{First eqs}b) gives $P_T=\mathcal{E}^{5/6}$. For later reference, we summarise the transition-region scalings:
\refstepcounter{equation}
$$
\label{First BD trans scalings} 
L_T=1, \quad T_T=1, \quad Q_T=\mathcal{E}^{5/6}, \quad P_T=\mathcal{E}^{5/6}.
\eqno{(\theequation \mathrm{a}\!\!-\!\!\mathrm{d})}
$$

Note that the pressure in the transition region is small, whereby at order unity the pressure in the neck region attenuates rapidly  from its value \eqref{bubble pressure} in the bubble to zero. This is consistent with our tacit assumption that the force condition \eqref{First force} is dominated by the bubble region. Furthermore, substituting the scalings \eqref{First BD trans scalings} into the flux-conservation equation (\ref{First eqs}a) confirms that vapour production is negligible in the transition region, whereby the order $\mathcal{E}^{5/6}$ flux density originates from the vapour production in the bubble and neck regions.

\subsection{Breakdown of first distinguished limit}\label{ssec:firstBD}
Recall that the above scalings, which imply the formation of a neck-bubble morphology for small $\mathcal{E}$, were derived from the leading-order thin-film problem of \S\ref{ssec:thinfilm}. The latter problem, in turn, was derived from the unified small-slope model by considering the limit $\mathcal{B}\to0$ with $\mathcal{E}\simeq1$. This suggests returning to the unified small-slope model to check, based on the scalings derived, whether new physics not included in the thin-film problem of \S\ref{ssec:thinfilm} become important for sufficiently small $\mathcal{E}$. 

\begin{figure}[htbp]
\begin{center}
\includegraphics[scale=0.37]{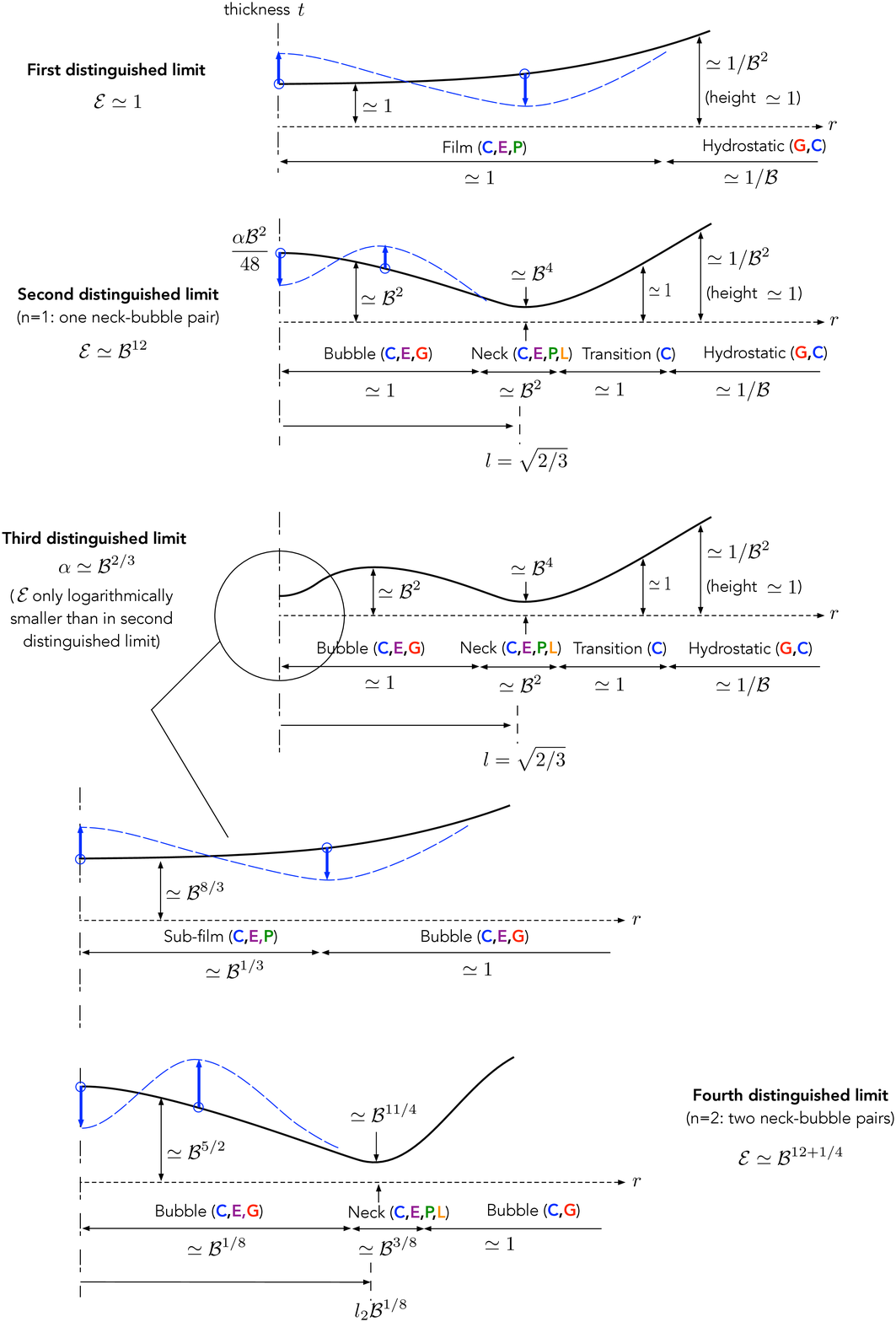}
\caption{Depiction of the first four distinguished limits (see \S\ref{sec:formation} and \S\ref{sec:onset}). The bold letters C, P, E and G indicate the importance of capillarity, pressure-driven flow, vapour production and variations in gravitational hydrostatic pressure, respectively, for determining the leading-order thickness profile; furthermore, bold $L$ indicates a radially localised region. The blue arrows and dashed curves indicate evolution of the film morphology with decreasing $\mathcal{E}$.}
\label{fig:lim1234}
\end{center}
\end{figure}

In particular, consider the dynamical condition \eqref{p t eq} written in terms of the thickness profile $t$. In the bubble region, the small-$\mathcal{E}$ scalings imply a dominant balance at order $\mathcal{E}^{1/6}$ which includes a uniform pressure correction and the curvature of the thickness profile. Given the smallness in $\mathcal{E}$ of these leading terms, it is plausible  that the gravity terms in \eqref{p t eq}, neglected in the leading-order thin-film problem on account of their smallness in $\mathcal{B}$, enter this dominant balance for sufficiently small $\mathcal{E}$. Indeed, using the scaling $L_B=1$ and the fact that $\hat{h}_{\infty}\simeq 1$ as $\mathcal{E}\to0$, we have the estimate $\mathcal{B}^2(h_{\infty}-r^2/2)\simeq \mathcal{B}^2$; since $T_B\ll1$, the remaining gravity term $\mathcal{B}^2 t$ is relatively negligible. Thus, the order $\mathcal{B}^{2}$ gravity terms in \eqref{p t eq} are not negligible 
in the order $\mathcal{E}^{1/6}$ balance of the dynamic condition 
for 
\begin{equation}\label{Limit2}
\mathcal{E}\simeq \mathcal{B}^{12}.
\end{equation}
For such small $\mathcal{E}$, the thin-film problem of \S\ref{ssec:thinfilm} no longer holds --- 
the first distinguished limit of the unified small-slope model breaks down. This process is schematically shown in Fig.~\ref{fig:lim1234}. We note that a similar breakdown, wherein gravity terms become important in the vapour film, does not occur in the case of small Leidenfrost drops levitating above a flat solid substrate.

\section{Onset of gravity effects in the thin-film region}\label{sec:onset}
\subsection{Second distinguished limit}\label{ssec:second}
We are led to analyse a second distinguished limit of the unified small-slope model, where both $\mathcal{B}$ and $\mathcal{E}$ are small in accordance with \eqref{Limit2}. We expect the thin-film region to decompose into bubble, neck and transition regions as described in the preceding section, only with gravity now having a leading-order effect in the bubble region. Accordingly, we expect the small-$\mathcal{E}$ scalings derived in \S\ref{ssec:neckbubble}, based on the first limit, to remain valid in the second. Furthermore, their analysis in the second limit should overlap with the small-$\mathcal{E}$ behaviour of the first limit, which is why we did not go beyond a scaling analysis in \S\ref{ssec:neckbubble}.

In analysing the second distinguished limit, it will be convenient to work with the single small parameter $\mathcal{B}$. We accordingly define the rescaled evaporation number
\begin{equation}\label{tilde E}
\mathcal{\tilde{E}}= \mathcal{E}/\mathcal{B}^{12}.
\end{equation}
Similarly, we use \eqref{Limit2} to rewrite the scalings derived in \S\ref{ssec:neckbubble} in terms of $\mathcal{B}$. Thus, the bubble scalings become
\refstepcounter{equation}
$$
\label{Second bubble scalings}
L_B=1, \quad T_B=\mathcal{B}^2, \quad Q_B=\mathcal{B}^{10},  \quad P_B'=\mathcal{B}^2, \quad \Delta P_B=\mathcal{B}^4,
\eqno{(\theequation \mathrm{a}\!\!-\!\!\mathrm{e})}
$$
the neck scalings become
\refstepcounter{equation}
$$
\label{Second neck scalings}
L_N = \mathcal{B}^2, \quad T_N = \mathcal{B}^4, \quad Q_N=\mathcal{B}^{10}
\eqno{(\theequation \mathrm{a}\!\!-\!\!\mathrm{c})}
$$
and the transition scalings become
\refstepcounter{equation}
$$
\label{Second trans scalings} 
L_T=1, \quad T_T=1, \quad Q_T=\mathcal{B}^{10}, \quad P_T=\mathcal{B}^{10}.
\eqno{(\theequation \mathrm{a}\!\!-\!\!\mathrm{d})}
$$

\subsubsection{Bubble region}
We start with the bubble region, where the scalings \eqref{Second bubble scalings} suggest the expansions
\refstepcounter{equation}
$$
\label{Second bubble expansions}
t= \mathcal{B}^{2}\tilde{t}(r) + \cdots, \quad q = \mathcal{B}^{10}\tilde{q}(r) + \cdots,\quad p =2+ \mathcal{B}^{2}\tilde{p} + \cdots,
\eqno{(\theequation \mathrm{a}\!\!-\!\!\mathrm{c})}
$$
where $\tilde{p}$ is constant and $0\le r < l$ (cf.~\eqref{bubble extent}). Expanding the sphere displacement as in (\ref{hydro h exp}b), the leading non-trivial balance of the dynamical condition \eqref{p t eq} yields
\begin{equation}\label{tilde t eq}
\frac{1}{r}\frac{d}{dr}\left(r\frac{d\tilde{t}}{dr}\right)=\hat{h}_{\infty}-\frac{1}{2}r^2-\tilde{p},
\end{equation}
which can be considered an equation for $\tilde{t}$. The first two terms on the right-hand side together account for the effect of gravity owing to the displacement between the local film height and the unperturbed bath. 
This equation is supplemented by the boundary conditions
\refstepcounter{equation}
$$
\label{tilde t bcs}
\frac{d\tilde{t}}{dr}=0 \quad \text{at} \quad r=0,  \qquad \tilde{t}=0 \quad \text{at} \quad r=l,
\eqno{(\theequation \mathrm{a},\!\mathrm{b})}
$$
where the first follows from (\ref{bcs at zero}a) and \eqref{t def}, while the second represents matching with the relatively thin neck region. Solving \eqref{tilde t eq} together with \eqref{tilde t bcs} gives 
\begin{equation}\label{tilde t sol}
\tilde{t}=\frac{1}{32}\left(l^2-r^2\right)\left(r^2+\alpha\right),
\end{equation}
in which
\begin{equation}\label{alpha def}
\alpha = l^2+8\left(\tilde{p}-\hat{h}_{\infty}\right)
\end{equation}
is a constant to be determined together with $\tilde{p}$ and $\hat{h}_{\infty}$. Note that $\alpha$ must be positive as otherwise $\tilde{t}$ would vanish within the bubble domain. 

Next, the leading-order balance of the flux-conservation equation (\ref{q eqs}a), together with \eqref{tilde E}, gives
\begin{equation}\label{tilde q eq}
\frac{1}{r}\frac{d}{dr}\left(r\tilde{q}\right)=\frac{\mathcal{\tilde{E}}}{\tilde{t}},
\end{equation}
to be solved for $\tilde{q}$ in conjunction with the boundary condition 
\begin{equation}\label{tilde q bc}
\tilde{q}=0 \quad \text{at} \quad r=0,
\end{equation}
which follows from (\ref{bcs at zero}b). 
Substituting \eqref{tilde t sol} and integrating using \eqref{tilde q bc} yields
\begin{equation}\label{tilde q sol}
\tilde{q} = \frac{16\mathcal{\tilde{E}}}{(l^2+\alpha) r}\ln \frac{(1+ r^2/\alpha)}{1-r^2/l^2}.
\end{equation}
For later reference, we note the behaviour
\begin{equation}\label{tilde q beh}
\tilde{q}\sim -\frac{8\sqrt{6}\mathcal{\tilde{E}}}{l^2+\alpha}\ln\frac{\sqrt{6}\left(l-r\right)\alpha}{l^2+\alpha} \quad \text{as} \quad r\nearrow l,
\end{equation}
where the asymptotic error is algebraic in the indicated limit.

\subsubsection{Neck region}\label{sssec:necksecond}
Consider next the neck region near $r=l$. With reference to the scalings \eqref{Second neck scalings}, we introduce the strained coordinate
\begin{equation}\label{R def}
R=(r-l)/\mathcal{B}^2
\end{equation}
and pose the expansions
\refstepcounter{equation}
$$
\label{Second neck expansions}
t=\mathcal{B}^4T(R)+\cdots, \quad q=\mathcal{B}^{10}Q(R)+\cdots, \quad p=P(R)+\cdots.
\eqno{(\theequation \mathrm{a}\!\!-\!\!\mathrm{c})}
$$
From \eqref{q eqs} and \eqref{p t eq}, we find the governing equations
\refstepcounter{equation}
$$
\label{Second neck eqs}
\frac{d{Q}}{dR}=\frac{\mathcal{\tilde{E}}}{{T}}, \quad {Q}=-\frac{{T}^3}{12}\frac{d{P}}{d R}, \quad 
{P}=2-\frac{d^2{T}}{d R^2}.
\eqno{(\theequation \mathrm{a}\!\!-\!\!\mathrm{c})}
$$
These are to be solved in conjunction with matching conditions as $R\to\pm\infty$ to be discussed below. 

It turns out that the parameter $\tilde{\mathcal{E}}$ can be factored out from both \eqref{Second neck eqs} and the matching conditions by use of the transformations
\refstepcounter{equation}
$$
\label{sim trans}
T(R)=\tilde{\mathcal{E}}^{1/3}T'(R'), \quad Q(R)=\tilde{\mathcal{E}}^{5/6}Q'(R'), \quad P(R)=P'(R'),
\eqno{(\theequation \mathrm{a}\!\!-\!\!\mathrm{c})}
$$
where
\begin{equation}\label{Rs def}
R=R_s+\tilde{\mathcal{E}}^{1/6}R'
\end{equation}
and $R_s$ is a constant shift discussed below.
In particular, \eqref{Second neck eqs} become
\refstepcounter{equation}
$$
\label{Second neck eqs sim}
\frac{dQ'}{dR'}=\frac{1}{T'}, \quad Q'=-\frac{T'^3}{12}\frac{dP'}{dR'}, \quad P'=2-\frac{d^2T'}{d R'^2}.
\eqno{(\theequation \mathrm{a}\!\!-\!\!\mathrm{c})}
$$

Consider now the conditions as $R\to\pm\infty$. Straightforward matching of the pressure field between the neck and bubble regions gives
\begin{equation}
\label{First neck Pp left}
P'\to2 \quad \text{as} \quad R'\to-\infty,
\end{equation}
while the asymptotically small magnitude of the pressure in the transition region implies 
\begin{equation}
\label{Second neck Pp left}
P'\to0 \quad \text{as} \quad R'\to\infty.
\end{equation}
With \eqref{First neck Pp left} and \eqref{Second neck Pp left}, local analysis of \eqref{Second neck eqs sim} as $R\to\pm\infty$ yields the behaviours
\refstepcounter{equation}
$$
\label{Second neck local left} 
T'\sim  -\theta R' , \quad Q'\sim  -\frac{1}{\theta}\ln  |R'| + \tau \quad \text{as} \quad R'\to-\infty
\eqno{(\theequation \mathrm{a},\!\mathrm{b})}
$$
and
\refstepcounter{equation}
$$
\label{Second neck local right} 
T'= R'^2 + O(1), \quad Q'\sim \nu  \quad \text{as} \quad R'\to\infty,
\eqno{(\theequation \mathrm{a},\!\mathrm{b})}
$$
where $\theta, \tau$ and $\nu$ are constants to be determined, and we have used the displacement invariance associated with the constant shift $R_s$ to eliminate the linear term in (\ref{Second neck local right}a); that choice defines the constant $R_s$, which we will calculate later. The transformed neck problem consisting of \eqref{Second neck eqs sim}, \eqref{Second neck local left} and \eqref{Second neck local right} determines two relations between the three unknowns $\theta, \tau$ and $\nu$, which we conceptually represent by functions: $\tau(\theta)$ and $\nu(\theta)$. 
In terms of these functions, asymptotic matching of the flux density between the bubble and neck regions provides the relations
\refstepcounter{equation}
$$
\label{trans canonical}
\theta=\frac{2+3\alpha}{24\sqrt{6}\tilde{\mathcal{E}}^{1/6}}, \quad  \theta \tau =\ln \frac{8 \theta}{\alpha \mathcal{B}^2}. 
\eqno{(\theequation \mathrm{a},\!\mathrm{b})}
$$
Numerical solution of the transformed neck problem together  with the conditions \eqref{trans canonical} determines $\alpha$ and $\theta$ (and hence also $\tau$ and $\nu$) as functions of $\tilde{\mathcal{E}}$ and $\ln \mathcal{B}$. 

As an alternative to the above numerical procedure,  further asymptotic reduction can be achieved by expanding the solutions in inverse logarithmic powers (recall our convention to group together terms of the same algebraic order). We do not pursue this here as the resulting approximations are  inaccurate while not adding significant insight.
\begin{figure}[t!]
\begin{center}
\includegraphics[scale=0.3]{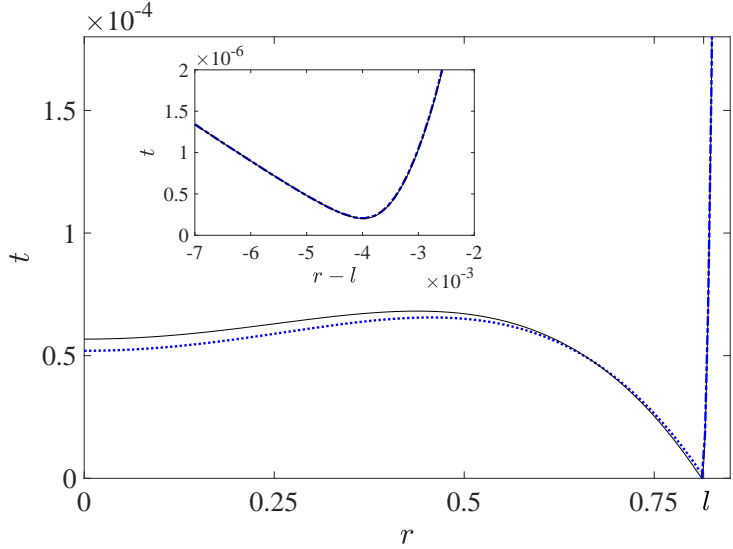}
\caption{The film-thickness profile for $\mathcal{B}=0.1$ and $\mathcal{E}=10^{-22}$. The numerical solution of the unified small-slope model is depicted by the black solid line. Asymptotic predictions in the second distinguished limit (\S\ref{ssec:second}) are depicted by the blue dotted line (bubble-region approximation) and dash-dotted line (neck-region approximation). The inset shows the vicinity of the neck near $r=l$.  }
\label{fig:2lim}
\end{center}
\end{figure}

\subsubsection{Transition and hydrostatic regions}\label{ssec:transition hydrostatic}
We continue with the transition region, where the scalings \eqref{First BD trans scalings} suggest the expansions 
\begin{equation}
h= \acute{h}(r)+\cdots, \quad t =\acute{t}(r) + \cdots,
\end{equation}
where $r>l$. From \eqref{t def}, the leading-order thickness and height profiles are related as
\begin{equation}
\acute{t}=\frac{1}{2}r^2-\acute{h}.
\end{equation}
Given the smallness of $p$ (cf.~(\ref{First BD trans scalings}d)), the dynamical condition \eqref{p h eq} gives
\begin{equation}\label{cute h eq}
\frac{1}{r}\frac{d}{dr}\left(r\frac{d\acute{h}}{dr}\right)=0.
\end{equation}
Boundary conditions on $\acute{h}$ can be derived by asymptotic matching between the neck and transition regions. Thus, matching of the thickness profile gives the conditions
\refstepcounter{equation}
$$
\label{transition bcs}
\acute{t}=0, \quad \frac{d\acute{t}}{dr}=0 \quad \text{at} \quad r=l,
\eqno{(\theequation \mathrm{a},\!\mathrm{b})}
$$
which are intuitive given the quadratic widening of the vapour film outward from the thin neck. In terms of $\acute{h}$, \eqref{transition bcs} read as
\refstepcounter{equation}
$$
\label{transition bcs height}
\acute{h}=\frac{1}{3}, \quad \frac{d\acute{h}}{dr}=\sqrt{\frac{2}{3}} \quad \text{at} \quad r=l.
\eqno{(\theequation \mathrm{a},\!\mathrm{b})}
$$
Solving \eqref{cute h eq} together with \eqref{transition bcs height}, we find
\begin{equation}\label{cute h}
\acute{h}=\frac{2}{3}\ln r + \frac{1}{3}\left(1+\ln\frac{3}{2}\right).
\end{equation}

The analysis of the hydrostatic region is similar to that in the first distinguished limit. In particular, the bath height $h$ is expanded as in (\ref{hydro h exp}a), with the leading-order field $\hat{h}$ satisfying the differential equation \eqref{phi eq} and far-field condition \eqref{phi far}. Together with asymptotic matching with the transition region (cf.~\eqref{cute h}), we find the same solution \eqref{hydro h} but with the leading-order sphere displacement given by  
\begin{equation}\label{Second varphi}
\hat{h}_{\infty} = \frac{2}{3}\ln\frac{
{\sqrt{6}}}{\mathcal{B}}+\frac{1}{3}-\frac{2}{3}\gamma_E.
\end{equation}

The value \eqref{Second varphi} is just the small-$\mathcal{E}$ limit \eqref{floating} of the formula \eqref{varphi} obtained for $\hat{h}_{\infty}$ in the analysis of the first distinguished limit. Conversely, the limit \eqref{beta E small} of the function $\beta$ appearing in that formula is readily deduced from  \eqref{cute h}. We have already noted that \eqref{floating} corresponds to the displacement of a non-wetting sphere floating at a liquid-bath interface. 

\subsubsection{Perturbed bubble pressure and shift of the neck radius}\label{ssec:perturbed and shift}
With $\hat{h}_{\infty}$ and $\alpha$ determined, the latter's definition \eqref{alpha def} yields the constant correction $\tilde{p}$ to the pressure in the bubble region as a function of $\mathcal{\tilde{E}}$ and $\ln\mathcal{B}$. Furthermore, carefully enforcing the force constraint \eqref{force} beyond leading order allows calculating the constant $R_s$, which corresponds to a small shift of the neck radius (cf.~\eqref{Rs def}). The details are given in Appendix \ref{App:Force}, where we find
\begin{equation}\label{Rs text}
R_s=\frac{1-6\hat{h}_{\infty}}{12\sqrt{6}}
\end{equation}
with $\hat{h}_{\infty}$ provided by \eqref{Second varphi}. While the value of $R_s$ does not affect the leading-order approximations obtained to this point, knowledge of it is necessary in order to compare our asymptotic approximations for the neck region with numerical solutions of the unified small-slope model. 

In Fig.~\ref{fig:2lim}, we compare the neck and bubble approximation obtained in the second distinguished limit with a numerical solution of the unified small-slope model for $\mathcal{B}=0.1$ and $\mathcal{E}=10^{-22}$. In plotting the neck profile we use the result \eqref{Rs text} for the displacement of the neck position from $r=l$. Furthermore, the approximation for the film thickness at the origin is depicted in Fig.~\ref{fig:thick} as a function of $\mathcal{E}$ by the dashed curve.

\subsection{Breakdown of second distinguished limit}\label{ssec:secondBD}
\subsubsection{Variation of bubble morphology with $\tilde{\mathcal{E}}$}
From the above calculation scheme, we numerically observe that $\alpha$ decreases monotonically with $\mathcal{\tilde{E}}$. Consider the corresponding change in the morphology of the bubble region, which is described by \eqref{tilde t sol}. For $\alpha > l^2$, the thickness profile has a local maximum at $r=0$ given by
\begin{equation}\label{tilde t min}
\tilde{t}(0)=\frac{\alpha}{48}.
\end{equation}
For $\alpha < l^2$, however, the thickness profile has a local minimum at $r=0$, again given by \eqref{tilde t min}, and an interior local maximum at $r=(l^2-\alpha)^{1/2}/2^{1/2}$ given by $(l^2+\alpha)/128$. 

Thus, as $\tilde{\mathcal{E}}$ is decreased, the bubble region transitions from a dimple-like morphology where the thickness is maximal at the origin, to an annular morphology where the  thickness is maximal at a positive radius smaller than the neck radius; following this transition, the latter maximal thickness continues to increase while the minimal thickness at the origin continues to decrease. This process is schematically shown in  Fig.~\ref{fig:lim1234}. Note that the variation with $\mathcal{E}$ of the thickness at the origin follows an opposite trend relative to that predicted in the first distinguished limit (\S\ref{ssec:neckbubble}).

In what follows, we look to the neck region to extract the asymptotic dependence of $\alpha$ upon $\tilde{\mathcal{E}}$. This will allow us to identify the breakdown of the second distinguished limit as the thickness at the origin becomes so small that the approximations underlying the description of 
the bubble region collapse. 

\subsubsection{Limit neck problem}\label{ssec:limitneck}
Consider the small $\tilde{\mathcal{E}}$ limit of the neck problem formulated in \S\ref{ssec:second}. From \eqref{trans canonical}, we have that 
\begin{equation}\label{theta E}
\theta \sim \frac{1}{12\sqrt{6}}\tilde{\mathcal{E}}^{-1/6} \quad \text{as} \quad \tilde{\mathcal{E}}\to 0. 
\end{equation}
We will use this  behaviour together with 
relation (\ref{trans canonical}b) to extract the corresponding  behaviour of $\alpha$.

Note that the relation (\ref{trans canonical}b) involves the function $\tau(\theta)$, which is governed by the transformed neck problem \eqref{Second neck eqs sim}--\eqref{Second neck local right}. Thus, an intermediate step is to study that problem as $\theta\to\infty$ in order to extract the behaviour of $\tau(\theta)$ in that limit. As a first step, a dominant-balance argument hints to the leading-order expansions 
\refstepcounter{equation}
$$
\label{dprime exp}
T'(R')\sim \theta^2T''(R''), \quad Q'(R')\sim \theta^5 Q''(R''), \quad P'(R') \sim P''(R''),
\eqno{(\theequation \mathrm{a}\!\!-\!\!\mathrm{c})}
$$
as $\theta\to\infty$, 
where we define the double-primed strained coordinate
\begin{equation}\label{dprime R}
R''=R'/\theta.
\end{equation}
Using these scalings, we see from the flux-conservation equation (\ref{Second neck eqs sim}a)  that vapour production is negligible at leading order, whereby 
\begin{equation}
Q''(R'')\equiv \chi,
\end{equation}
where $\chi$ is a constant; from \eqref{Second neck local left} and \eqref{Second neck local right}, the functions $\tau$ and $\nu$ are then seen to possess the leading-order behaviours
\refstepcounter{equation}
$$
\label{tau nu asym}
\tau \sim \chi \theta^5, \quad \nu \sim \chi \theta^5,
\eqno{(\theequation \mathrm{a},\!\mathrm{b})}
$$
as $\theta\to\infty$. With $Q''$ being constant, the neck problem reduces to the differential equation 
\begin{equation}\label{dprimed eq}
\frac{T''^3}{12}\frac{d^3T''}{d{R''}^3}=\chi,
\end{equation}
which is obtained by combining (\ref{Second neck eqs sim}a) and (\ref{Second neck eqs sim}b), together with the far-field conditions
\refstepcounter{equation}
$$
\label{dprimed far fields}
T''\sim -{R''}   \quad \text{as} \quad R''\to - \infty, \quad 
T''= {R''}^2 + O(1) \quad \text{as} \quad R''\to\infty,
\eqno{(\theequation \mathrm{a},\!\mathrm{b})}
$$
which follow from \eqref{Second neck local left} and \eqref{Second neck local right}, respectively. The differential equation \eqref{dprimed eq} is third order, while the far-field expansions \eqref{dprimed far fields} effectively provide four auxiliary conditions. The extra condition serves to determine the constant $\chi$.

The constant-flux neck problem consisting of \eqref{dprimed eq} and \eqref{dprimed far fields} is essentially equivalent to that considered in Appendix D of  \cite{Jones:78}. Adopting the numerical result given there to our notation, we have
\begin{equation}\label{chi value}
\chi \doteq 0.051.
\end{equation}
In Appendix \ref{App:CanNeckNum}, we formulate a slightly generalised constant-flux neck problem which can be reduced to the present one via straightforward rescaling. It will be convenient to refer to this generalised problem later in the paper.

Combining (\ref{tau nu asym}a) and \eqref{chi value}, we have a leading-order approximation for $\tau(\theta)$ as $\theta\to\infty$. It turns out, however, that in order to extract from (\ref{trans canonical}b) the small $\tilde{\mathcal{E}}$ behaviour of $\alpha$, we actually need $\tau$ to higher order. To that end, we substitute the far-field behaviour (\ref{dprimed far fields}a) into the leading nontrivial balance of the primed flux-conservation equation (\ref{Second neck eqs sim}a), written in terms of the double-primed coordinate  and thickness profile; comparison with the primed far-field condition (\ref{Second neck local left}b), which defines $\tau$, then gives
\begin{equation}\label{tau large theta}
\tau \sim \chi \theta^5+\frac{\ln \theta-\ln k}{\theta} \quad \text{as} \quad \theta\to\infty,
\end{equation}
where $k$ is a pure constant whose actual value is not important for our purposes. ($k$ can be determined by continuing the analysis of the neck problem to higher orders as $\theta\to\infty$.)  

Using \eqref{theta E} and \eqref{tau large theta}, we find from (\ref{trans canonical}b) the behaviour
\begin{equation}\label{log alpha asymptotics}
\ln \frac{1}{\alpha} =  \frac{\chi}{(12\sqrt{6})^6}\frac{1}{\tilde{\mathcal{E}}}+\ln\frac{\mathcal{B}^2}{8} + \ln\frac{1}{k} +o(1) \quad \text{as} \quad \tilde{\mathcal{E}}\to0. 
\end{equation}
It is this behaviour of the logarithm of $\alpha$ that will be needed later, rather than the behaviour of $\alpha$ itself. For the sake of completeness, we also give the latter as
\begin{equation}\label{alpha asymptotics}
\alpha \sim \frac{8k}{\mathcal{B}^2}\exp\left\{{-\frac{\chi}{(12\sqrt{6})^6}\frac{1}{\tilde{\mathcal{E}}}}\right\} \quad \text{as} \quad \tilde{\mathcal{E}}\to0,
\end{equation}
showing more explicitly that $\alpha$ vanishes exponentially as $\tilde{\mathcal{E}}\to0$. Note that terms up to order unity in \eqref{log alpha asymptotics} are crucial to determining the leading-order behaviour \eqref{alpha asymptotics}.

\subsubsection{Formation of sub-film region and breakdown of second distinguished limit}
From \eqref{tilde t min}, the rescaled bubble thickness at the origin vanishes linearly as $\alpha\to0$. Accordingly, the quadratic behaviour of the rescaled thickness profile $\tilde{t}(r)$ (cf.~ \eqref{tilde t sol}) hints to the formation of a new sub-film region near the origin where $r\simeq \alpha^{1/2}$ and $\tilde{t}\simeq\alpha$. Furthermore, \eqref{tilde q sol} and \eqref{log alpha asymptotics} together imply that $\tilde{q}\simeq 1/\alpha^{1/2}$ in that region. Using \eqref{Second bubble expansions}, we see that in terms of the original dimensionless variables introduced in \S\ref{ssec:nondim}, the sub-film radial extent, thickness and flux density respectively possess the scalings
\refstepcounter{equation}
$$
\label{film scalings}
L_S=\alpha^{1/2}, \quad T_S=\mathcal{B}^2\alpha, \quad Q_S=\mathcal{B}^{10}\alpha^{-1/2}.
\eqno{(\theequation \mathrm{a}\!\!-\!\!\mathrm{c})} 
$$

The above scalings are based on the asymptotic model derived in the second distinguished limit. Returning to the original pressure-flux relation  (\ref{q eqs}b), however, we find using \eqref{film scalings} that the pressure variation across the sub-film region scales like
\begin{equation}\label{film variation scaling}
\Delta P_S 
=\frac{\mathcal{B}^4}{\alpha^3}.
\end{equation}
There is therefore the possibility that for $\alpha$ sufficiently small pressure variations may no longer be small compared to the constant pressure correction $\mathcal{B}^2\tilde{p}$ appearing in the pressure expansion (\ref{Second bubble expansions}c). Indeed, it follows from \eqref{alpha def} and \eqref{Second varphi} that $\tilde{p}\simeq 1$ as $\alpha\to0$. Accordingly, the second distinguished limit breaks down for 
\begin{equation}\label{Limit3}
\alpha \simeq \mathcal{B}^{2/3}.
\end{equation}

With \eqref{alpha asymptotics},  condition \eqref{Limit3} implies that the second distinguished limit breaks down when $\tilde{\mathcal{E}}$ is just logarithmically small, i.e., of order $1/\ln \mathcal{B}$. Continuing to group  
terms that are of the same algebraic order in $\mathcal{B}$, this breakdown is seen to occur at the same algebraic scaling \eqref{Limit2} of $\mathcal{E}$ with $\mathcal{B}$ which was used to define the second distinguished limit in the first place.   
Nonetheless, the exponential behaviour of $\alpha$ as $\tilde{\mathcal{E}}\to0$, as expressed by \eqref{alpha asymptotics}, provides us with the  magnifying glass required to identify the breakdown of the second distinguished limit and next to continue our analysis to even smaller $\mathcal{E}$, beyond the domain of validity of the approximation obtained in that limit. 

\subsection{Third distinguished limit}\label{ssec:third}
Guided by the above discussion, we shall investigate a third distinguished limit  where the algebraic scaling \eqref{Limit2} of $\mathcal{E}$ with $\mathcal{B}$ still holds despite $\alpha$ being small as in \eqref{Limit3}; as shown schematically in Fig.~\ref{fig:lim1234}, we expect a sub-film region near the origin which is asymptotically distinct from the bubble region. With \eqref{Limit2} still valid, the asymptotic analyses in \S\ref{ssec:second} of the bubble, neck, transition and hydrostatic regions need to be only slightly modified. Below, we outline these modifications and subsequently analyse the new sub-film region. 

\subsubsection{Transition and hydrostatic regions}
We first briefly consider the transition and hydrostatic regions. It is readily seen that the leading-order solutions \eqref{hydro h} and \eqref{cute h}, respectively describing the hydrostatic and transition regions, remain valid in the third distinguished limit with the leading-order sphere displacement given by \eqref{Second varphi}. Indeed, these solutions were determined based on the matching conditions \eqref{transition bcs} which simply follow from the leading-order position of the neck region; since we still expect the pressure to be approximately $2$ in the bubble region (as well as in the smaller sub-film region), the force argument given above \eqref{bubble extent}  still holds. 

\subsubsection{Bubble region}\label{3vnt}
Consider next the bubble region, with the fields expanded as in \S\ref{ssec:second}.  The leading-order film thickness $\tilde{t}$ is still governed by the differential equation \eqref{tilde t eq} and boundary condition (\ref{tilde t bcs}b). In contrast, the symmetry boundary condition (\ref{tilde t bcs}a) no longer holds; instead, matching with the sub-film region implies 
\begin{equation}\label{third tilde t matching}
\tilde{t}=0 \quad \text{at} \quad r=0,
\end{equation}
since the radial extent and thickness of the sub-film  region are both small compared to the corresponding properties of the bubble region. Solving \eqref{tilde t eq} together with (\ref{tilde t bcs}b) and \eqref{third tilde t matching} yields
\begin{equation}\label{tilde t 3}
\tilde{t}=\frac{1}{32}\left(l^2-r^2\right) r^2,
\end{equation}
which as expected is the same as \eqref{tilde t sol} for $\alpha=0$. It follows that expression \eqref{alpha def} for the constant pressure correction $\tilde{p}$ reduces to 
\begin{equation}\label{modified tildeP}
\tilde{p} = \hat{h}_{\infty} - \frac{1}{12}.
\end{equation}

Following the analysis of the bubble region in \S\ref{ssec:second}, we next consider the leading-order flux density $\tilde{q}$. Thus, integrating \eqref{tilde q eq}, with $\tilde{t}$ given by \eqref{tilde t 3}, we find
\begin{equation}\label{Q2}
\tilde{q} = \frac{24\mathcal{\tilde{E}}}{r}\left(\ln \frac{l^2r^2}{l^2-r^2}+\zeta\right),
\end{equation}
where $\zeta$ is a constant to be determined. (Note that the symmetry condition \eqref{tilde q bc} no longer applies.) For later reference, we note the behaviours 
\refstepcounter{equation}
$$
\tilde{q}\sim 12\sqrt{6}\tilde{\mathcal{E}}\left\{\ln \frac{l}{3\left(l-r\right)}+\zeta\right\} \; \text{as} \quad r\to l, \quad 
\tilde{q} \sim \frac{24\tilde{\mathcal{E}}}{r}\left(2\ln r+\zeta\right) \quad \text{as} \quad r\to0,
\eqno{(\theequation \mathrm{a},\!\mathrm{b})}
$$
where for both expansions the error is algebraic in the indicated limit. 

Comparing with the analysis in \S\ref{ssec:second}, we see that the constant $\zeta$ effectively replaces $\alpha$ as the single unknown  appearing in the leading-order description of the bubble region. Analogously with that analysis, we expect that $\zeta$ could be determined via matching with the neck region; the value of $\zeta$ should then influence the solution in the sub-film region via matching. Thus, in the present regime, information propagates in the inward radial direction. 

\subsubsection{Neck region}
The leading-order problem in the neck region is essentially the same as in \S\ref{ssec:second}, only that the two matching conditions \eqref{trans canonical} need to be updated based on revised matching of flux between the bubble and neck regions. First, instead of (\ref{trans canonical}a), we find
\begin{equation}\label{Third theta}
 \theta = \frac{1}{12\sqrt{6}\tilde{\mathcal{E}}^{1/6}},
 \end{equation}
which is just (\ref{trans canonical}a) for $\alpha=0$. Unlike (\ref{trans canonical}a),  \eqref{Third theta} gives $\theta$ directly in terms of $\tilde{\mathcal{E}}$, thence also the function $\tau(\theta)$ defined by solution of the transformed neck problem (cf.~\eqref{Second neck eqs sim}--\eqref{Second neck local right}). Second, instead of  (\ref{trans canonical}b), we find
\begin{equation}\label{zeta def}
\zeta = \theta \tau - \ln \frac{8\theta}{\mathcal{B}^2},
\end{equation}
which determines $\zeta$ in terms of $\mathcal{\tilde{E}}$ and $\ln \mathcal{B}$. 

\subsubsection{Sub-film region}\label{sssec:subfilm}
Consider now the sub-film region near the origin. The scalings \eqref{film scalings} and \eqref{film variation scaling}, which characterise the sub-film region, can be expressed solely in terms of $\mathcal{B}$ by substituting the condition \eqref{Limit3}, which defines the third distinguished limit. This suggests defining the strained coordinate 
\begin{equation}
\check{r} = r/\mathcal{B}^{1/3}
\end{equation}
and posing the expansions
\refstepcounter{equation}
$$
 t= \mathcal{B}^{8/3}\check{t}+ \cdots, \quad q=\mathcal{B}^{29/3}\check{q}+ \cdots,\quad p =2+ \mathcal{B}^{2}\check{p} + \cdots.
\eqno{(\theequation \mathrm{a}\!\!-\!\!\mathrm{c})}
$$
The sub-film problem then consists of the differential equations (cf.~\eqref{q eqs} and \eqref{p t eq}) 
\refstepcounter{equation}
$$
\label{3ode}
\frac{1}{\check{r}}\frac{d}{d \check{r}}\left(\check{r}\check{q}\right)=\frac{\mathcal{\tilde{E}}}{\check{t}}, \quad \check{q}=-\frac{\check{t}^3}{12}\frac{d\check{p}}{d \check{r}}, \quad  \check{p}=  \hat{h}_{\infty}-\frac{1}{\check{r}}\frac{d}{d\check{r}}\left(\check{r}\frac{d \check{t}}{d\check{r}}\right),
\eqno{(\theequation \mathrm{a}\!\!-\!\!\mathrm{c})}
$$
the symmetry boundary conditions (cf.~\eqref{bcs at zero})
\refstepcounter{equation}
$$
\frac{d\check{t}}{d\check{r}}=0, \quad\check{q}=0 \quad \text{at} \quad \check{r}=0,
\eqno{(\theequation \mathrm{a},\!\mathrm{b})}
$$
and the far-field conditions
\refstepcounter{equation}
$$
\label{check far}
\check{t} \sim \frac{1}{48}\check{r}^2, 
\quad 
\check{q} \sim \frac{48\mathcal{\tilde{E}}}{\check{r}} \ln \check{r} + \frac{24\tilde{\mathcal{E}}}{\check{r}}\left(l^2\ln \mathcal{B}+\zeta\right) \quad \text{as} \quad \check{r}\to\infty,
\eqno{(\theequation \mathrm{a},\!\mathrm{b})}
$$
which follow from matching the film thickness and flux density between the sub-film and bubble regions. Note that the leading logarithmic term in (\ref{check far}b) is coupled to the leading quadratic term in (\ref{check far}a). Thus, it is the constant term in (\ref{check far}b) which serves as the fourth auxiliary boundary condition that closes the above fourth-order problem. In line with our earlier comment regarding information propagating radially inwards, the sub-film problem does not involve any undetermined constants. 

In Fig.~\ref{fig:thick}, the film thickness at the origin, as predicted by the above sub-film problem, is depicted by the dash-dotted line. (As explained below, the sub-film problem can be solved using the scheme devised for solving the film problem formulated in \S\ref{ssec:thinfilm}.) The agreement with the numerical solution of the unified small-slope model is noticeably not as good as in the first and second distinguished limits, though this is to be expected (for the moderately small value of $\mathcal{B}$ used in the numerical simulations) given that the scalings characterising the second and third limits are very close. Nonetheless, the asymptotic approximation obtained in the third distinguished limit qualitatively agrees with the numerical data.

\subsection{Second neck-bubble pair and breakdown of third distinguished limit}\label{ssec:thirdBD}
The sub-film problem is mathematically similar to the film problem found in the first distinguished limit (see \S\ref{ssec:thinfilm}), 
despite these regions being characterised by markedly differing scales and the matching here being with the bubble region rather than directly with the hydrostatic region. 
In fact, we show in Appendix \ref{App:ThirdBreak} that, despite the different form of far-field conditions,  the sub-film problem can be exactly mapped to the film problem in \S\ref{ssec:thinfilm}. The mapping is nontrivial, however, in the sense that it is defined by stretching factors that are governed by transcendental equations. Using that mapping, we show in Appendix \ref{App:ThirdBreak} that the behaviour of the solutions to the sub-film problem as $\mathcal{\tilde{E}}\to0$ is analogous to the behaviour discussed in \S\ref{ssec:firstBD} of the solutions to the film problem as $\mathcal{E}\to0$. 

The mathematical analogy with the film problem in the first distinguished limit implies that, as $\tilde{\mathcal{E}}$ is decreased, $\check{t}(\check{r})$ transitions from having a local maximum to a local minimum at the origin. Moreover, $\check{t}(0)$ vanishes as $\tilde{\mathcal{E}}\to0$, leading to the creation of a neck-bubble pair near the origin. Combining the mapping developed in Appendix \ref{App:ThirdBreak}, the scalings characterising the the sub-film and film regions and the discussion in \S\ref{ssec:firstBD} of the creation of the original neck-bubble pair prior to the breakdown of the first distinguished limit, we find that the new bubble region is characterised by the scalings
\refstepcounter{equation}
$$
\label{new bubble scalings}
L_{B,2}=\mathcal{B}^{1/3}\tilde{\mathcal{E}}^{-5/6}, \quad T_{B,2}=\mathcal{B}^{8/3}\tilde{\mathcal{E}}^{-2/3}, \quad Q_{B,2}=\mathcal{B}^{29/3}\tilde{\mathcal{E}}^{5/6}
\eqno{(\theequation \mathrm{a}\!\!-\!\!\mathrm{c})}
$$
while the new neck region is characterised by the scalings 
\refstepcounter{equation}
$$
\label{new neck scalings}
L_{N,2}=\mathcal{B}^{1/3}\tilde{\mathcal{E}}^{1/6}, \quad T_{N,2}=\mathcal{B}^{8/3}\tilde{\mathcal{E}}^{1/3}, \quad Q_{N,2}=\mathcal{B}^{29/3}\tilde{\mathcal{E}}^{5/6},
\eqno{(\theequation \mathrm{a}\!\!-\!\!\mathrm{c})}
$$
where as usual $L$, $T$ and $Q$ refer to radial extent, thickness and flux density. These scalings show that the new bubble region is asymptotically smaller than the original one, both in thickness and radial extent (as long as $\tilde{\mathcal{E}} \gg \mathcal{B}^{2/5}$, but this condition turns out to be irrelevant given the breakdown discussed below).

Analogously to the breakdown of the first distinguished limit, we find that variations in gravitational hydrostatic pressure in the unified small-slope model, negligible in the sub-film region of the third distinguished limit, become non-negligible in the new bubble region for sufficiently small $\tilde{\mathcal{E}}$. With reference to the interfacial dynamical condition \eqref{p t eq}, this occurs when the order $T_{B,2}/L_{B,2}^2$ curvature of the thickness profile becomes as small as the order $\mathcal{B}^2 L_{B,2}^2$ hydrostatic pressure associated with the quadratic variation in the height of the sphere. It follows that the third distinguished limit breaks down for 
\begin{equation}\label{Limit4}
\tilde{\mathcal{E}}\simeq \mathcal{B}^{1/4} \Rightarrow \mathcal{E}\simeq \mathcal{B}^{12+1/4}.
\end{equation}
Thus, there is a fourth distinguished limit where $\mathcal{E}$ scales as in \eqref{Limit4}. Substituting \eqref{Limit4} into the scalings \eqref{new bubble scalings} and \eqref{new neck scalings}, and referring back to the the unified small-slope model, we see that in this new distinguished limit vapour production is non-negligible only  in the new neck-bubble pair. See Fig.~\ref{fig:lim1234} for a schematic depiction of the fourth distinguished limit.

\section{Formation of capillary waves}\label{sec:wave formation}
In contrast with the clear separation between the first and second limits, the scalings characterising the second, third and fourth limits are very close; moreover, the scalings characterising the secondary neck-bubble pair associated with the latter two limits are near unity. Indeed, this new feature of the morphology is barely visible numerically and so 
 pragmatically the third and fourth limits are of little interest. It is evident, however, that the asymptotic evolution of the thin-film morphology does not terminate with the fourth limit. Numerically, we observe that continuing to decrease $\mathcal{E}$ (for a fixed small value of $\mathcal{B}$) brings out completely new features in the thin-film morphology that do not correspond to any of the limiting descriptions already identified; in particular, we observe the formation of a region near the origin of nearly uniform thickness, which subsequently grows radially into the main bubble region. This motivates us to continue tracking the asymptotic evolution of the thin film to smaller $\mathcal{E}$, in order to identify those distinguished limits which are, in fact, pragmatically significant to describing the thin-film morphology. 

Towards unravelling the next stage of the evolution of the thin-film morphology, we hypothesise that the process by which neck-bubble pairs are created near the origin essentially repeats itself; in particular, we anticipate a sequence of analogues to the second distinguished limit, where the $n$th analogue features a chain of $n$ neck-bubble pairs, all but the first of which localised near the origin.  In the following subsection \S\ref{ssec:scaling analysis}, we use scaling argument to study this scenario; in \S\ref{ssec:seqlimits}, we identify a  distinguished limit corresponding to each value of $n$; and in \S\S\ref{ssec:seqBD}--\ref{ssec:form uniform} we describe how the sequence of distinguished limits eventually terminates, leading to a new stage in the evolution of the thin-film morphology. Detailed analysis for arbitrary $n$, going beyond scaling arguments, is provided in Appendix \ref{App:NBubbles}.

\subsection{Chains of neck-bubble pairs}\label{ssec:scaling analysis}
Consider the scenario hypothesised above, where the asymptotic structure of the thin film includes $n$ pairs of neck and bubble regions. We use the index $m=1,2,\ldots,n$ to identify the pairs, starting from the outermost pair adjacent to the transition region and ending at  the innermost one closest to the origin. Let  $L_{N,m}$, $T_{N,m}$ and $Q_{N,m}$, respectively, be the scalings of the radial extent, thickness and flux density for the $m$th neck, and $L_{B,m}$, $T_{B,m}$ and $Q_{B,m}$ the corresponding scalings for the $m$th bubble. In determining these scalings, we shall specifically assume a scenario analogous to the second (and fourth) distinguished limits, wherein hydrostatic-pressure variations are important in all of the bubble regions; we shall accordingly be skipping the corresponding analogues of the first (and third) distinguished limits. This assumption effectively slaves $\mathcal{E}$ to $\mathcal{B}$, whereby the requisite scalings will be determined solely in terms of $\mathcal{B}$.

Based on what we have seen in the first four distinguished limits, we anticipate that:
\begin{enumerate}
\item The thickness and radial extent of the bubbles attenuate inwards.
\item Successive bubbles are separated by a relatively narrow neck, i.e., $L_{N,m} \ll L_{B,m}$. It follows that 
\begin{equation}\label{neck position scaling}
\text{radial position of } m\text{th neck } \simeq  L_{B,m}.
\end{equation}
\item The radial extent of the first bubble region remains the same when there are multiple neck-bubble pairs:
 \begin{equation}\label{first length}
L_{B,1}=1.
\end{equation}
In particular, this was the case for the secondary neck-bubble pair that emerged in the third and fourth distinguished limits (the latter corresponding to `$n=2$' here). 

Alternatively, we could assume that the pressure is order unity across all neck and bubble regions. Together with the first and last of this list, the scaling \eqref{first length} would in that case follow from a leading-order balance of the force constraint \eqref{force}. 
\item The scalings characterising the transition and hydrostatic regions remain the same as in the previous distinguished limits. 
\end{enumerate}

In the neck regions, the pressure variation associated with the capillarity term in \eqref{p t eq} (i.e., the curvature of the thickness profile) drives a flux in accordance with the flux-pressure relation (\ref{q eqs}b). Thus, if $\Delta P_{N,m}$ denotes the scaling of the pressure variation across that neck, we find from \eqref{p t eq} and (\ref{q eqs}b) that 
\refstepcounter{equation}
$$
\label{DP neck scalings}
\Delta P_{N,m}= \frac{T_{N,m}}{L_{N,m}^2}, \quad Q_{N,m}=T_{N,m}^3\frac{\Delta P_{N,m}}{L_{N,m}}.
\eqno{(\theequation{\mathrm{a},\!\mathrm{b}})}
$$
These two relations can be combined into
\begin{equation}\label{r1}
T_{N,m}^4 = Q_{N,m}L_{N,m}^3.
\end{equation}
In the bubble regions, the capillarity term and the quadratic hydrostatic term in \eqref{p t eq} are expected to be comparable (analogously to the case in the second distinguished limit), i.e., 
\begin{equation}\label{r2}
T_{B,m} = \mathcal{B}^2 L_{B,m}^4.
\end{equation}

To relate the scalings that describe successive regions, we derive further scaling relations  based on asymptotic matching. 
We verified in \S\ref{ssec:second} that, for the first neck region, the thickness profile increases quadratically to the right (cf.~(\ref{Second neck local right}a)) and linearly to the left (cf.~(\ref{Second neck local left}a)).
For a chain of neck-bubble pairs where the bubbles attenuate inwards, we expect this to be the case for all the necks. (This can be understood \textit{a posteriori}  from the discussion below regarding the pressure field.) Accordingly, the leading-order thickness profile for each bubble region vanishes quadratically towards the origin and linearly towards its outward bounding neck. (The quadratic vanishing at the origin holds both for bubbles $m=1,\ldots,n-1$, on which scale the position of the inward bounding neck is indistinguishable from the origin, and the final $n$th bubble, where axial symmetry applies at the origin.) Matching  therefore implies the relations
\refstepcounter{equation}
$$
\label{r3}
 \frac{T_{B,m}}{L_{B,m}^2} =   \frac{T_{N,m+1}}{L_{N,m+1}^2},\qquad  \frac{T_{B,m}}{L_{B,m}} =   \frac{T_{N,m}}{L_{N,m}}.
\eqno{(\theequation{\mathrm{a},\!\mathrm{b}})}
$$

We next relate the scalings of the flux density between the different regions. In regions where the vapour-production term in (cf.~\ref{q eqs}a) can be neglected, continuity dictates that $q$ attenuates like $1/r$. In that case, \eqref{neck position scaling} and $L_{B,m+1}\ll L_{B,m}$ imply the relations
\refstepcounter{equation}
$$
\label{r4}
Q_{B,m} L_{B,m} =  Q_{B,m-1} L_{B,m-1}, \quad Q_{B,m}=Q_{N,m}.
\eqno{(\theequation{\mathrm{a},\!\mathrm{b}})}
$$
A more detailed scaling argument shows that \eqref{r4} also hold for regions where vapour production is non-negligible  \footnote{Consider a given neck region where vapour production is non-negligible in the flux-conservation equation (\ref{q eqs}a). Performing a local analysis of that equation, using the previously noted quadratic vs.~linear growth of the thickness profile away from the neck, shows that the flux density in the neck approaches a constant value towards its outward neighbouring bubble and increases logarithmically towards its inward neighbouring bubble (e.g., see (\ref{Second neck local left}b) and (\ref{Second neck local right}b).) It follows from matching principles that \eqref{r4} holds also for the value of $m$ corresponding to the neck considered here. (We will see that the vapour production term in (\ref{q eqs}a) is important only for the innermost neck, whereby that value of $m$ is necessarily $n$.)}.

Solving the recurrence relations \eqref{r1}--\eqref{r4} together with the initial condition $L_{B,1}=1$, we find 
\begin{subequations}
  \label{scalings sm}
\begin{gather}
L_{B,m} = \mathcal{B}^{s_m}, \quad T_{B,m} = \mathcal{B}^{2 + 4 s_m},\quad Q_{B,m}=\mathcal{B}^{10-s_m},\quad \nonumber \tag{\theequation\!a-\!-\!-\!c}\\ 
\qquad L_{N,m} =  \mathcal{B}^{2 - 13 s_m},\quad  T_{N,m} =  \mathcal{B}^{4 - 10 s_m}, \quad Q_{N,m} =  \mathcal{B}^{10-s_m}, \tag{\theequation\!d-\!-\!-\!f}
\end{gather}
 \end{subequations}
wherein
\begin{equation}\label{sm}
s_m=\frac{8}{7}\left(\frac{1}{8}-\frac{1}{8^m}\right), \quad m=1,2,\ldots,n,
\end{equation}
is a monotonically increasing sequence that approaches $1/7$ as $m\to\infty$ (assuming $n$ can be arbitrarily large, see \S\ref{ssec:seqBD}).
Note that the scalings \eqref{scalings sm} are consistent with those obtained in \S\ref{sec:onset} for the single neck-bubble pair in the second distinguished limit and the two neck-bubble pairs in the fourth distinguished limit.

We can check that the scalings \eqref{scalings sm} are consistent with the assumptions listed at the beginning of this subsection. This is evident for the first two assumptions. Next, substituting \eqref{scalings sm}  into the dynamic condition \eqref{p t eq} shows that $p\simeq 1$ across all neck-bubble pairs, therefore our third assumption is also consistent. In fact, the same argument suggests that $p\sim 2$ in all of those regions, excluding the first neck region across which the pressure at order unity must drop to zero; thus, the radial extent of the first bubble remains $\sim l$ (cf.~\eqref{neck position scaling}). The fourth assumption is evident given that  the scalings of the first neck-bubble pair remain as before.

In those regions where $p\sim 2$, the leading-order film thickness and flux density are influenced by higher-order terms in an expansion for $p$.  This motivates us to introduce a reduced pressure $p'=p-2-\mathcal{B}^2 h_{\infty}$, which the dynamic condition \eqref{p t eq} gives as
\begin{equation}\label{reduced pressure}
p'=-\frac{\mathcal{B}^2}{2}r^2+\mathcal{B}^2t-\frac{1}{r}\frac{d}{dr}\left(r\frac{dt}{dr}\right).
\end{equation}
The reduced pressure includes only non-uniform contributions. In particular, the scalings \eqref{scalings sm} show that the second term in \eqref{reduced pressure} is negligible compared to the first (quadratic-hydrostatic)  and last (capillarity) contributions for all neck and bubble regions. Furthermore, let $P'_{N,m}$ and $P'_{B,m}$ be the scalings of the reduced pressure in the $m$th neck and bubble, respectively. For the neck regions, the capillarity contribution in \eqref{reduced pressure} dominates the reduced pressure, hence  \eqref{scalings sm} gives
\begin{equation}\label{neck pressure scalings}
P'_{N,m} =\mathcal{B}^{16s_m}.
\end{equation}  
From \eqref{DP neck scalings}, this is also the scaling  $\Delta P_{N,m}$ of the pressure variation across the $m$th neck. 
For the bubble regions, the by-design balance between the capillarity and quadratic-hydrostatic contributions in the dynamic condition \eqref{p t eq}  implies that these terms together dominate the reduced pressure; it therefore follows from \eqref{scalings sm} and \eqref{reduced pressure} that $P'_{B,m}=\mathcal{B}^2L_{B,m}^2$. The flux-pressure equation (\ref{q eqs}b), however, suggests a generally different scaling $\Delta P_{B,n}=Q_{B,m}L_{B,m}/T_{B,m}^3$ for the pressure variation. Thus, using \eqref{scalings sm}, we find that
\begin{equation}\label{bubble pressure scalings}
P'_{B,m}=\mathcal{B}^{2+2s_m}\quad\gg\quad\Delta P_{B,m}=\mathcal{B}^{4-12s_m}.
\end{equation}  
We see that the reduced pressure in the bubble regions is uniform, to leading order, consistently with the second distinguished limit (cf.~(\ref{Second bubble expansions}c)). From these comments we can deduce that, in the bubble regions, the pressure possesses the  expansions 
\begin{subequations}
  \label{P expansion bubble}
\begin{gather}
p= 2 + \mathcal{B}^2\tilde{p}_1+\cdots, \quad \text{for} \quad m=1,\nonumber  \tag{\theequation\!a}\\ 
p= 2 + \mathcal{B}^2\hat{h}_{\infty}+\mathcal{B}^{2+2s_m}\tilde{p}_m +\cdots, \quad \text{for} \quad m=2,3,\ldots,n,\nonumber   \tag{\theequation\!b}
\end{gather} 
 \end{subequations}
wherein $\tilde{p}_m$ are constants; similarly, for the neck regions,
  \begin{subequations}
  \label{P expansion neck}
\begin{gather}
p= P_1 +\cdots, \quad \text{for} \quad m=1,\nonumber  \tag{\theequation\!a}\\
p= 2+\mathcal{B}^2P_2 +\cdots, \quad \text{for} \quad m=2,\nonumber  \tag{\theequation\!b}\\
p= 2+\mathcal{B}^2\hat{h}_{\infty}+\mathcal{B}^{16s_m}P_m +\cdots, \quad \text{for} \quad m=3,4,\ldots,n,\nonumber  \tag{\theequation\!c}
\end{gather}
 \end{subequations}
wherein $P_m$ are functions of the corresponding neck coordinates $(r-L_{B,m}l_m)/L_{N,m}$,  $L_{B,m}l_m$ being the approximate position of the $m$th neck (in particular, $l_1=l$). In \eqref{P expansion bubble} and \eqref{P expansion neck}, we have approximated the sphere deflection $h_{\infty}$ by its leading-order value $\hat{h}_{\infty}$ \eqref{varphi}; it can be shown that higher-order corrections do not affect the above expansions up to the orders shown \footnote{Since $s_m$ is monotonically increasing and $s_m\to1/7$ as $m\to\infty$, it is sufficient to show that $\mathcal{B}^2(h_{\infty} -\hat{h}_{\infty}) \ll \mathcal{B}^{2 + 2/7}$ or equivalently that $h_{\infty} = \hat{h}_{\infty} + o(\mathcal{B}^{2/7})$. This can be shown by solving the problem for the height profile in the transition and hydrostatic regions and matching the solutions with the leading-order solution in the outermost neck region.}.

\subsection{Sequence of distinguished limits}\label{ssec:seqlimits}
In the preceding scaling analysis, the number $n$ of neck-bubble pairs appears to be arbitrary. We now show that $n$ is effectively determined by the relative smallness of $\mathcal{E}$ and $\mathcal{B}$. To that end, consider the innermost (i.e., $n$th) bubble region, which is unique in that axial symmetry applies and dictates that the flux density $q$ must vanish at the origin (cf.~(\ref{bcs at zero}b)). For that to be possible, however, the left- and right-hand sides of the flux-conservation equation $(\ref{q eqs}a)$ must balance to avoid a leading-order solution that is singular at the origin. With reference to the scalings \eqref{scalings sm}, that balance implies a  sequence of distinguished limits:
\begin{equation}\label{Blimit}
\mathcal{E} \simeq \mathcal{B}^{12 + 2 s_n},\quad n=1,2,\ldots,
\end{equation}
respectively associated with a vapour-film morphology exhibiting a chain of $n$ neck-bubble pairs as described in \S\ref{ssec:scaling analysis}. 

\begin{figure}[b!]
\begin{center}
\includegraphics[scale=0.33]{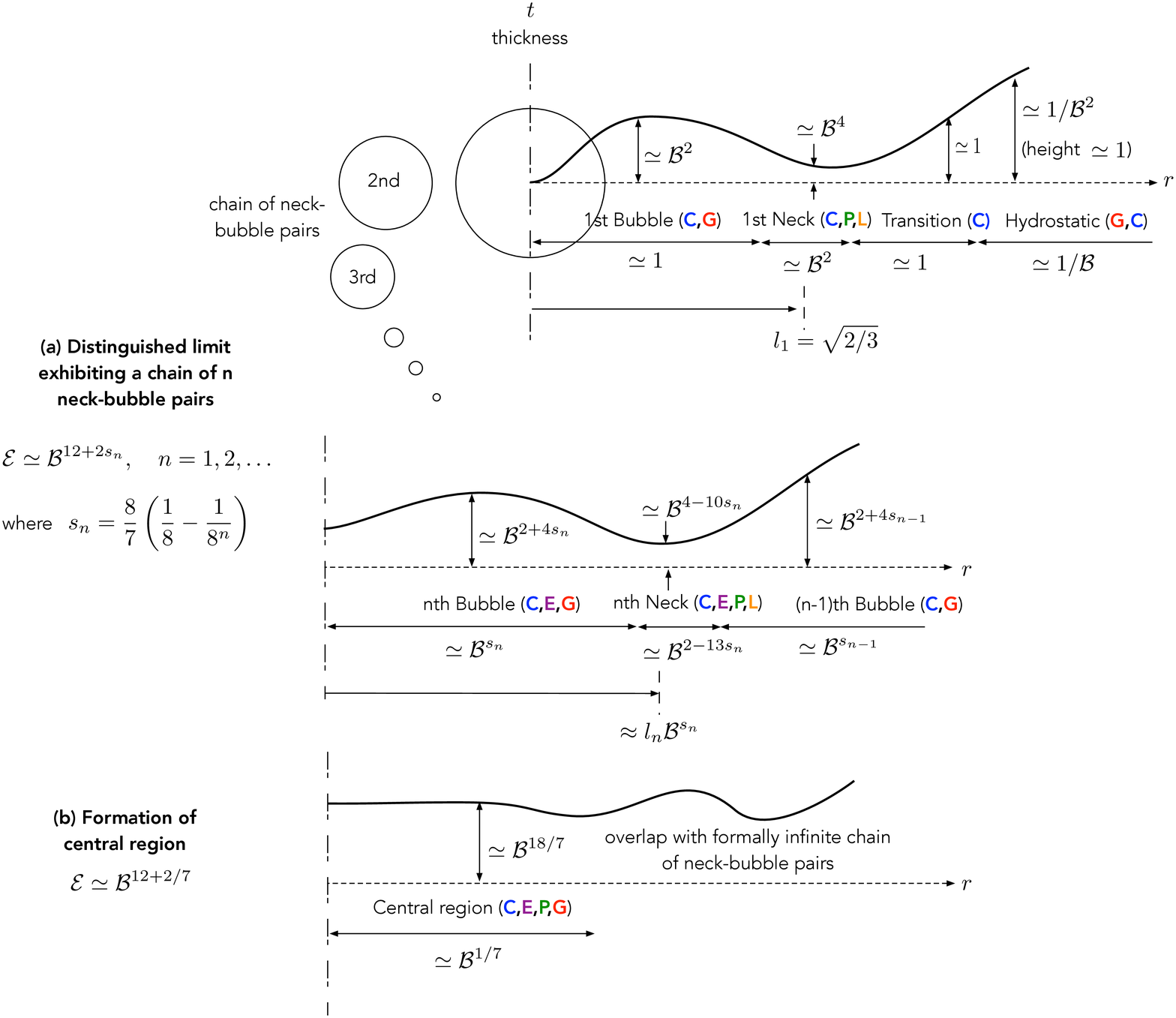}
\caption{Depiction of the $n$th element of the sequence \eqref{Blimit} of distinguished limits (\textbf{a}) and the distinguished limit \eqref{LimitInf} corresponding to the accumulation point of that sequence (\textbf{b}). The bold letters C, P, E, G and L have the same meaning as in Fig.~\ref{fig:lim1234}.}
\label{fig:1seq}
\end{center}
\end{figure}

As expected, \eqref{Blimit} reduces to  the second distinguished limit \eqref{Limit2} for $n=1$, and the fourth distinguished limit \eqref{Limit4} for $n=2$.  Note that substitution of \eqref{scalings sm} and  \eqref{Blimit}  into the flux-conservation equation (\ref{q eqs}a) shows that the vapour-production term therein is negligible for all neck-bubble pairs but the innermost one, consistently with what we found in the second and fourth distinguished limits. 
A schematic of the distinguished limit corresponding to the $n$th term in the sequence \eqref{Blimit} is depicted in Fig.~\ref{fig:1seq}a. 

In order to further corroborate the above picture, in Appendix \ref{App:NBubbles} we build on the scalings derived in this section to construct a detailed asymptotic description of the vapour film in the distinguished limit corresponding to the $n$th term in the sequence \eqref{Blimit}. We find that, for each $n$, the solution in each neck-bubble pair can be recursively determined starting from the outermost pair ($m=1$).  Thus, the solution for pair $m$ is independent of that for pair $m+1$ and subsequent pairs. In particular, the flux density in pairs $m<n$, where vapour production is negligible, can be determined from continuity, in conjunction with the knowledge of the flux density through the outermost neck; we thereby find in Appendix \ref{App:NBubbles} that the approximation
\begin{equation}\label{cont flux}
q \sim \mathcal{B}^{10} \frac{\chi l}{(12\sqrt{6})^5}\frac{1}{r}
\end{equation}
holds, for $n>1$, everywhere except in the innermost pair (cf.~\eqref{NBchirel}). 

\subsection{Accumulation of distinguished limits and formation of central region}\label{ssec:seqBD}
Formally, taking the limit as $n\to\infty$ in \eqref{Blimit} gives the scaling
\begin{equation}\label{LimitInf}
\mathcal{E}\simeq B^{12+2/7}.
\end{equation}
In that limit, the scalings \eqref{scalings sm} and \eqref{neck pressure scalings}--\eqref{bubble pressure scalings} of the extent, thickness and flux density for  the innermost neck-bubble pair give
\refstepcounter{equation}
$$
\label{LimitScal}
L_{B,n},L_{N,n} \to \mathcal{B}^{1/7},\quad T_{B,n},T_{N,n} \to \mathcal{B}^{18/7},\quad Q_{B,n},Q_{N,n} \to  \mathcal{B}^{69/7},
\eqno{(\theequation{\mathrm{a}\!\!-\!\!\mathrm{c}})}
$$
while the corresponding scalings of the pressure variation and reduced pressure give
\begin{equation}
\label{LimitPressure}
\Delta P_{B,n},\Delta P_{N,n},P'_{B,n},P'_{N,m}\to\mathcal{B}^{16/7}. 
\end{equation}

The present picture, however, where the vapour film features an increasing number $n$ of neck-bubble pairs, namely gravity-capillary waves, as $\mathcal{E}$ is decreased or $\mathcal{B}$ is increased, does not hold for arbitrarily large $n$. Indeed, large $n$ implies that there are numerous neck-bubble pairs localised near the origin. But for $m$ large like $\ln\ln 1/\mathcal{B}$, scrutiny of \eqref{scalings sm} shows that the corresponding neck and bubble regions are, in fact, characterised by the same algebraic scalings; additionaly, neighbouring neck-bubble pairs are characterised by the same algebraic scalings. Thus, for any fixed value of $\mathcal{B}$, no matter how small, the maximal number of neck-bubble pairs formed as $\mathcal{E}$ is decreased is not infinite, but rather large like $\ln\ln1/\mathcal{B}$. Incidentally, the latter very mild scaling rationalises why numerically we observe at most a few oscillations of the film's thickness profile. 

We shall refer to the scaling \eqref{LimitInf} as the `accumulation point' of the sequence \eqref{Blimit} of distinguished limits, since the limits defined by that sequence which are associated with $n$ of order $\ln\ln1/\mathcal{B}$ and larger are effectively inseparable from the accumulation-point limit \eqref{LimitInf}. In that limit, we expect the picture of a chain of neck-bubble regions terminating at an innermost bubble to fail sufficiently close to the origin; we instead hypothesise an outer, formally infinite, chain of neck-bubble regions that matches a new central region localised at the origin (as depicted schematically in Fig.~\ref{fig:1seq}b). From \eqref{LimitScal} and \eqref{LimitPressure} we deduce the scalings 
\refstepcounter{equation}
$$
\label{central scalings}
L_{C} = \mathcal{B}^{1/7},\quad T_{C} = \mathcal{B}^{18/7},\quad Q_{C} =  \mathcal{B}^{69/7},\quad P_C',\Delta P_{C} =  \mathcal{B}^{16/7},
\eqno{(\theequation{\mathrm{a}\!\!-\!\!\mathrm{d}})}
$$
of the radial extent, thickness, flux, reduced pressure and pressure variation, respectively, which characterise that central region. 

Recall that the finite chains of neck-bubble pairs of \S\ref{ssec:scaling analysis} and \S\ref{ssec:seqlimits} were determined via a recursive process in which information propagates inwards. Accordingly, in the accumulation-point limit \eqref{LimitInf} we expect the formally infinite chain of neck-bubble regions to retain the same scalings as derived in \S\ref{ssec:scaling analysis} (and asymptotic description as derived in  Appendix \ref{App:NBubbles}) for finite chains, only that now there is no innermost pair and vapour production is negligible for all neck-bubble pairs.  Indeed, substitution of the scalings \eqref{scalings sm} and \eqref{central scalings} into (\ref{q eqs}a) confirms that vapour production is appreciable only in the central region.

\subsection{Central region}\label{ssec:central}
We now study the central region anticipated in the accumulation-point limit \eqref{LimitInf}. Assuming that limit, and given the scalings \eqref{central scalings}, we define the rescaled evaporation number
\begin{equation}\label{inf E}
\breve{\mathcal{E}}=\mathcal{E}/\mathcal{B}^{12+2/7},
\end{equation}
the strained radial coordinate
\begin{equation}
\breve{r}=r/\mathcal{B}^{1/7}
\end{equation}
and pose the expansions
\refstepcounter{equation}
$$
t= \mathcal{B}^{18/7}\breve{t}(\breve{r})+\cdots, \quad q= \mathcal{B}^{69/7}\breve{q}(\breve{r}) +\cdots, \quad p = 2 + \mathcal{B}^2\hat{h}_{\infty} + \mathcal{B}^{16/7}\breve{p}(\breve{r}) +\cdots,
\eqno{(\theequation{\mathrm{a}\!\!-\!\mathrm{c}})}
$$
where we have approximated the sphere displacement $h_{\infty}$ by its leading-order value $\hat{h}_{\infty}$, for the same reasons noted below expansions \eqref{P expansion bubble} and \eqref{P expansion neck}. 

From \eqref{q eqs} and \eqref{p t eq}, the leading-order fields satisfy  
\refstepcounter{equation}
$$
\label{eqiw}
\frac{1}{\breve{r}}\frac{ d (\breve{r}\breve{q})}{d \breve{r}}=\frac{\breve{\mathcal{E}}}{\breve{t}}, \quad \breve{q}=-\frac{\breve{t}^3}{12}\frac{d\breve{p}}{d \breve{r}}, \quad \breve{p} = -\frac{1}{2}\breve{r}^2 -\frac{1}{\breve{r}}\frac{d}{d\breve{r}}\left(\breve{r}\frac{d\breve{t}}{d\breve{r}}\right),
\eqno{(\theequation{\mathrm{a}\!\!-\!\!\mathrm{c}})}
$$
while \eqref{bcs at zero} provide the symmetry conditions
\refstepcounter{equation}
$$
\label{bcs central}
\frac{d\breve{t}}{d\breve{r}}=0, \quad  \breve{q} = 0 \quad \text{at} \quad \breve{r}=0.
\eqno{(\theequation \mathrm{a},\!\mathrm{b})}
$$

Note that the governing equations \eqref{eqiw} involve all the physical mechanisms present in the neck and bubble regions described in \S\ref{ssec:scaling analysis} and \S\ref{ssec:seqlimits}, including vapour production, which there was non-negligible only for the innermost pair. This is perhaps unsurprising given that the central region arises from the the accumulation of indistinguishable neck and bubble regions localised near the origin. 

The above formulation of the problem governing the central region is missing far-field conditions as $\breve{r}\to\infty$, which should follow from asymptotic matching between the central region and the formally infinite chain of neck-bubble pairs. One such condition, 
\begin{equation}\label{q far}
\breve{q}\sim \frac{\chi l}{(12\sqrt{6})^5}\frac{1}{\breve{r}},
\end{equation}
readily follows from approximation \eqref{cont flux} for the flux density, which holds throughout the neck-bubble chain given that the vapour-production term in (\ref{q eqs}a) is only non-negligible in the central region. Given \eqref{q far}, for large $\breve{r}$ we find that (\ref{eqiw}b) and (\ref{eqiw}c) reduce to 
\refstepcounter{equation}
$$
\label{eq far}
\frac{\chi l}{(12\sqrt{6})^5}\frac{1}{\breve{r}}\sim-\frac{\breve{t}^3}{12}\frac{d\breve{p}}{d\breve{r}} , \quad \breve{p}\sim-\frac{1}{2}\breve{r}^2-\frac{1}{\breve{r}}\frac{d}{d\breve{r}}\left(\breve{r}\frac{d\breve{t}}{d\breve{r}}\right).
\eqno{(\theequation \mathrm{a},\!\mathrm{b})}
$$
These equations contain precisely the leading-order physics governing the neck-bubble chain, including both the neck and bubble regions. They therefore constitute a `composite' approximation that could, in principle, be used to describe that chain as a whole (as an alternative to the matched asymptotics analysis in Appendix \ref{App:NBubbles}). 

This suggests that the central region and neck-bubble chain asymptotically overlap and can therefore be systematically matched. Traditional matching principles \cite{Hinch:book}, however, are tailored to matching asymptotically distinguished regions. They  are therefore inadequate in the present scenario, where the matching is between a distinguished central region and a region comprised of a formally infinite chain of distinguished neck and bubble regions. Such an unconventional relation between asymptotic regions has been encountered in many other problems involving  capillary waves \cite{Jones:78,Wilson:83,Bowles:95,Jensen:97,Duchemin:05,Benilov:08,Benilov:10,Cuesta:12,Cuesta:14,Benilov:15,Hewitt:15,Van:19}. To the best of our knowledge, detailed matching has never been systematically attempted in these situations, although \textit{ad hoc}  numerical schemes have been devised \cite{Jones:78}. 

In the present problem, we envisage this matching could be carried out by first using \eqref{eq far} to describe the large-$\breve{r}$ behaviours of the central-region fields in terms of a formally infinite neck-bubble chain, and subsequently comparing the latter chain and the actual one, whose detailed asymptotic description is derived in Appendix \ref{App:NBubbles}. We assume that such a comparison would allow extracting sufficient information pertaining to  the large-$\breve{r}$ chain in order to formulate a numerical scheme for solving the central-region problem. Given the two boundary conditions \eqref{bcs central} and the far-field condition \eqref{q far}, only one additional independent condition would be needed. 

We shall not attempt this matching procedure here, partly because a detailed solution of the central region is of little interest and partly because the development of new asymptotic-matching techniques lies beyond the scope of this paper. Instead, we shall exploit the formulation of the central-region problem to extend our asymptotic description of the vapour-film morphology beyond the distinguished scaling \eqref{LimitInf}, i.e., to $\mathcal{E}\ll \mathcal{B}^{12+2/7}$.

\subsection{Formation of a uniform-film region}\label{ssec:form uniform}
Consider the behaviour of the solution to the central-region problem for small $\breve{\mathcal{E}}$. Notwithstanding the latter smallness, there must still be a region containing the origin wherein the vapour-production term in (\ref{q eqs}a) remains non-negligible. Otherwise, the leading-order flux density would be singular at the origin. The vapour produced in this region should, moreover, be compatible with the flux density demanded by \eqref{q far}. A dominant-balance argument of \eqref{eqiw} then shows that, to leading order as $\breve{\mathcal{E}}\to0$, the dynamic condition (\ref{eqiw}c) degenerates to a balance between pressure and gravity,
\begin{equation}\label{p uniform film}
\breve{p}\sim -\frac{1}{2}{\breve{r}}^2, 
\end{equation}
whereas (\ref{eqiw}a) and (\ref{eqiw}b) remain valid in that limit. Solving the resulting leading-order problem in conjunction with the boundary condition \eqref{bcs central}, we find the explicit approximations 
\refstepcounter{equation}
$$
\label{breve sol}
\breve{t} \sim \left(6 \breve{\mathcal{E}}\right)^{1/4},\quad \breve{q} \sim  \frac{1}{12}\left(6
\breve{\mathcal{E}}\right)^{3/4}\breve{r},\eqno{(\theequation{\mathrm{a},\!\mathrm{b}})}
$$
representing a region containing the origin which consists of a film of uniform thickness, with the flux density growing in proportion to the radius. Furthermore, since vapour production is supposed to be negligible except in this uniform-film region, comparison of (\ref{breve sol}b) with the demanded flux density \eqref{q far} suggests that the latter region terminates abruptly at a value of $\breve{r}$ given by 
\begin{equation}\label{rf}
\breve{r}_f= \frac{1}{(12\sqrt{6})^2} \sqrt{\frac{\chi}{3}}\left(6 \breve{\mathcal{E}}\right)^{-3/8},
\end{equation}
in a manner which will be clarified in a more general context in the next section. 

Note that $\breve{r}_f$ is asymptotically large as $\breve{\mathcal{E}}\to0$. Hence, the central region \emph{expands} in that limit, with its innermost part morphing into a uniform-thickness film. For later reference, we give the scalings characterising this uniform-film region in terms of $\mathcal{B}$ and $\mathcal{E}$. Thus, using \eqref{central scalings}, \eqref{inf E} and \eqref{p uniform film}--\eqref{rf}, we find for the radial extent, thickness, flux density and pressure variation the respective scalings 
\refstepcounter{equation}
$$
\label{uniform scalings}
 L_U=\mathcal{B}^{19/4}\mathcal{E}^{-3/8}, \; T_U=\mathcal{B}^{-1/2}\mathcal{E}^{1/4}, \; Q_U=\mathcal{B}^{21/4}\mathcal{E}^{3/8},\;  \Delta P_{U} =  \mathcal{B}^{23/2}\mathcal{E}^{-3/4}.
\eqno{(\theequation{\mathrm{a}\!\!-\!\!\mathrm{d}})}
$$

\section{Expansion of uniform film into chain of neck-bubble pairs}\label{Sec:uniform1}
\subsection{Second formally infinite sequence of distinguished limits} \label{ssec:2seq}
In the previous section, we identified the formally infinite sequence \eqref{Blimit} of distinguished limits and subsequently the accumulation-point limit \eqref{LimitInf}. In the latter limit, we found a central region near the origin which  matches unconventionally with a formally infinite chain of neck-bubble regions. In the present section, we track the film morphology as  $\mathcal{E}$ is further decreased, or $\mathcal{B}$ is further increased. 

Our starting point is the analysis in \S\ref{ssec:form uniform} of the central-region problem as $\breve{\mathcal{E}}\to0$, which revealed that the inner part of that region expands into a uniform film. Eventually, the radial extent of the uniform film may become comparable to the radial extent of the $k$th $(k=1,2,\ldots)$ bubble in the neck-bubble chain, which order $L_{B,{k}}$ is given by (\ref{scalings sm}a). In that scenario, we expect the larger neck-bubble pairs ($m<k$) to remain intact, the $k$th bubble region to be reduced in size and modified by the expansion of the uniform-film region, and the smaller pairs ($m > k$) to be `squeezed' , in a sense to be described, into localised oscillations sandwiched between the uniform-film region and the modified $k$th bubble region.

The above discussion implies a second sequence of distinguished limits defined by the condition $L_U = L_{B,{k}}$. Using  (\ref{scalings sm}a) and (\ref{uniform scalings}a), the latter condition explicitly yields 
\begin{equation}\label{Blimit2}
\mathcal{E} \simeq \mathcal{B}^{12 + (2 -8 s_{{k}})/3},\quad k=1,2,\ldots,
\end{equation}
where $s_k$ is the sequence given by \eqref{sm}. The accumulation point \eqref{LimitInf} of the first sequence of distinguished limits \eqref{Blimit} is also the accumulation point of the second sequence of distinguished limits \eqref{Blimit2}. Here, however, decreasing $\mathcal{E}$ corresponds to the index $k$ decreasing from infinity to unity, rather than the other way around.  Analogously to the first sequence of distinguished limits, the limits represented by the terms of the sequence \eqref{Blimit2} are closely spaced; in practice, only the `last' limit corresponding to $k=1$ can be clearly observed in numerical solutions of the unified small-scale model.  

\begin{figure}[t!]
\begin{center}
\includegraphics[scale=0.29]{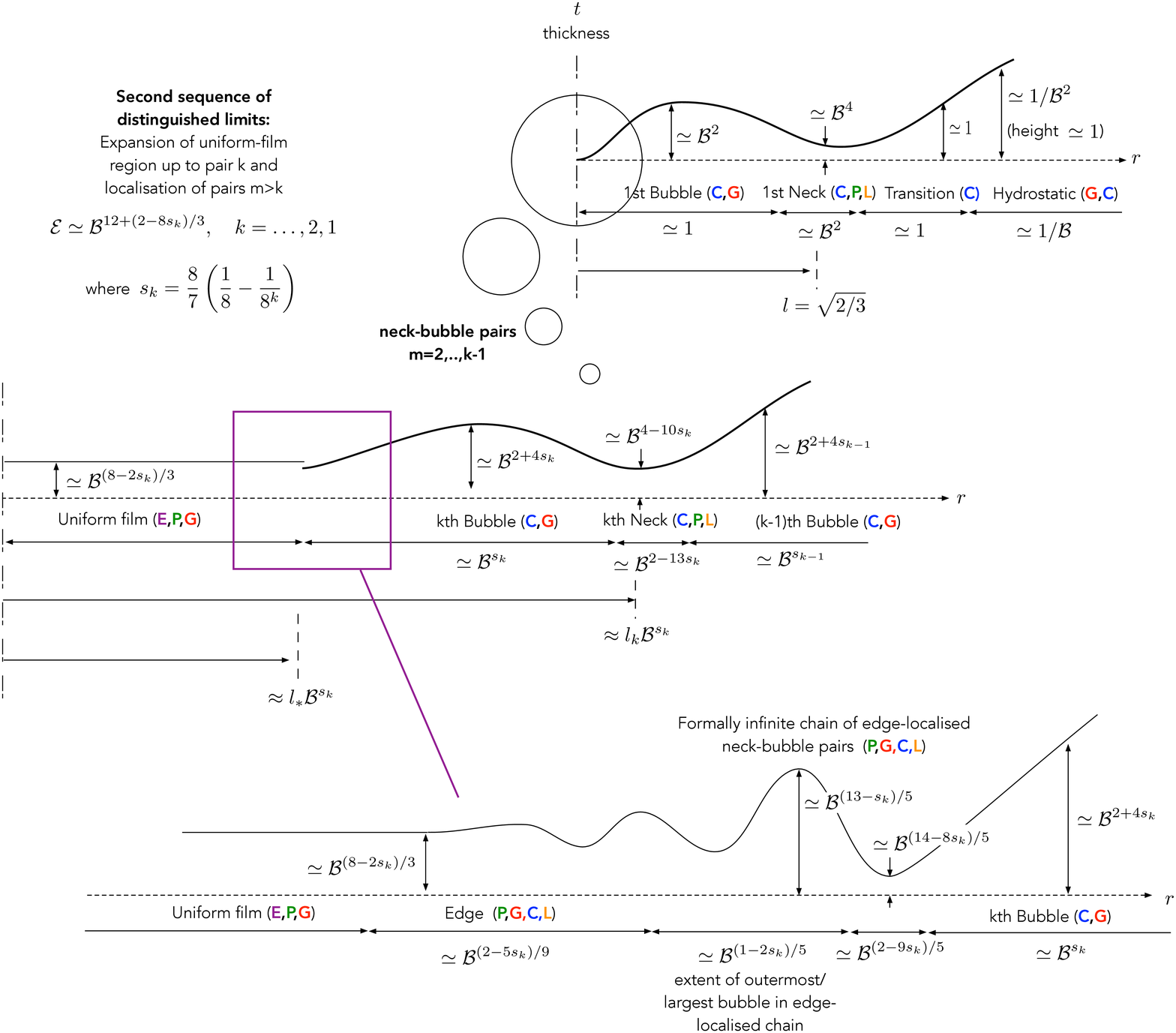}
\caption{Depiction of the $k$th element of the second sequence \eqref{Blimit2} of distinguished limits. The bold letters C, P, E, G and L have the same meaning as in Fig.~\ref{fig:lim1234}.}
\label{fig:2seq}
\end{center}
\end{figure}

In the remainder of this section,  we consider the distinguished limit corresponding to the $k$th term in \eqref{Blimit2}. It is convenient for that purpose to define the rescaled evaporation number
\begin{equation}\label{hat En}
\hat{\mathcal{E}}_k =\mathcal{E} /\mathcal{B}^{12 + (2 -8 s_{{k}})/3}.
\end{equation}
In that limit, we anticipate that the  vapour domain decomposes into the following asymptotic regions (schematically depicted in Fig.~\ref{fig:2seq}): 
\begin{enumerate}
\item A uniform-film region containing the origin, similar to the one discussed in \S\ref{ssec:form uniform}. This innermost region is the only region where the vapour-production term in (\ref{q eqs}a) is expected to be non-negligible. 
\item $k-1$ neck-bubble pairs whose leading-order description remains the same as pairs $m<n$ described in \S\ref{sec:wave formation} and Appendix \ref{App:NBubbles}, for which vapour production is negligible, with $n=k$ and the index $m =1,2,\ldots, k-1$  used as therein to count the pairs starting from the largest and outermost one. Note that these  bubble regions are large relative to the radial extent of the uniform-film region.
\item The $k$th neck-bubble pair. The radial extent of the $k$th bubble region is comparable to that of the uniform-film region; hence, the former region is expected to be substantially modified by the expansion of the uniform-film region.
\item A localised sequence of narrow edge regions, which as we will see is necessary to bridge the uniform-film region and the modified $k$th bubble region.
\item Transition and hydrostatic regions as in  \S\ref{ssec:transition hydrostatic}. 
\end{enumerate}

In this section our goal is to describe the above decomposition in more detail, stopping short of a detailed analysis of all regions. Thus, in \S\ref{ssec:uniform 2} we briefly revisit the uniform-film region and in \S\ref{ssec:edge} we qualitatively describe the edge regions alluded to above. In the subsequent section \S\ref{Sec:uniform2}, however, we carry out a detailed analysis of the last limit ($k=1$) and obtain accurate approximations describing the `squishing' of the outermost bubble by the expanding uniform-film region.

\subsection{Uniform film}\label{ssec:uniform 2}

The scalings characterising the uniform-film region in the present limit readily follow from substitution of \eqref{Blimit2} into \eqref{uniform scalings}:
\refstepcounter{equation}
$$
\label{uniform film scalings}
L_{U}=\mathcal{B}^{s_k}, \quad T_{U}=\mathcal{B}^{(8-2s_k)/3}, \quad Q_{U}=\mathcal{B}^{10-s_k},\quad \Delta P_{U} =  \mathcal{B}^{2 + 2 s_k}.
\eqno{(\theequation{\mathrm{a}\!\!-\!\!\mathrm{d}})}
$$
The analysis of this region closely follows that in \S\ref{ssec:form uniform}. Thus, we introduce the strained coordinate 
\begin{equation}\label{rk}
r_k =  r/\mathcal{B}^{s_k},
\end{equation}
pose the expansions
\begin{subequations}
\label{expansion uniform}
\begin{gather}
t= \mathcal{B}^{(8-2s_k)/3} \hat{t}(r_k)  \cdots, \quad q= \mathcal{B}^{10-s_k}\hat{q}(r_k) + \cdots, \nonumber \tag{\theequation\!$\mathrm{a}$,$\mathrm{b}$}\\ 
 p \sim 2 + \mathcal{B}^2\hat{h}_{\infty} +\mathcal{B}^{2+2s_k}\hat{p}(r_k) +\cdots\tag{\theequation\!c} 
\end{gather}
 \end{subequations}
and derived, from the leading-order balances of \eqref{q eqs}, \eqref{p t eq} and \eqref{bcs at zero}, the approximations
\refstepcounter{equation}
$$
\label{sol uniform}
\hat{t} =  \left(6\hat{\mathcal{E}}_k \right)^{1/4}  , \quad \hat{q} =  \frac{1}{12} \left(6 \hat{\mathcal{E}}_k \right)^{3/4} r_k, \quad \hat{p} =-\frac{1}{2}r_k^2.
\eqno{(\theequation{\mathrm{a}\!\!-\!\!\mathrm{c}})}
$$
(These are identical to the solutions  \eqref{p uniform film} and \eqref{breve sol} obtained in \S\ref{ssec:form uniform}.)

As in \S\ref{ssec:form uniform}, we expect the uniform region to be finite. We accordingly write 
\begin{equation}\label{uniform extent}
\text{extent of uniform film } \sim \mathcal{B}^{s_k} l_*,
\end{equation}
namely that the uniform-film region ends at $\hat{r}=l_*$, 
where $l_*$ depends solely on $\hat{\mathcal{E}}_k$. For $k>1$, the outermost neck-bubble pair remains unmodified and hence the approximation \eqref{cont flux} holds everywhere excluding the uniform-film region, where that result is perturbed by appreciable vapour production. Matching the latter approximation with (\ref{sol uniform}b) at the edge of the uniform film gives
\begin{equation}\label{l star}
 l_* = \frac{1}{(12\sqrt{6})^2} \sqrt{\frac{\chi}{3}}\left(6 \hat{\mathcal{E}}_k \right)^{-3/8}\quad \text{for} \quad k > 1,
\end{equation}
which is identical to the result \eqref{rf} already obtained in \S\ref{ssec:form uniform}. In the case $k=1$, in which the uniform-film region does modify the outermost neck-bubble pair, a more detailed analysis is necessary to determine $l_*$ (see \S\ref{Sec:uniform2}).

Recall Fig.~\ref{fig:thick}, where we plot the film thickness at the origin as a function of $\mathcal{E}$, comparing  numerical solutions of the unified small-slope model with asymptotic approximations derived in several  distinguished limits. In that figure, the present  approximation (\ref{sol uniform}a) for the thickness of the uniform-film region is depicted by the solid red line, showing excellent agreement over a wide range of $\mathcal{E}$. 

\subsection{Localised capillary waves at the edge of the uniform-film region}\label{ssec:edge}
Clearly, it is not possible to directly match the uniform-film region and the $k$th bubble region. This hints to the existence of an edge region localised near the outer rim of the uniform film. Let $L_E$, $T_E$, $Q_E$ and $\Delta P_E$ be the scalings of the extent, thickness, flux density and pressure variation in that region, respectively. Matching between the finite-extent uniform-film region and the edge region implies the scalings $T_E = T_U$ and $Q_E = Q_U$. Moreover, the pressure variation in the uniform film is dominated by the quadratic-hydrostatic gravitational contribution, whereas capillarity is also important in the $k$th bubble; this hints that both of these physical mechanisms should be present in the edge region. With reference to the uniform-film scalings \eqref{uniform film scalings} and the dynamic condition \eqref{p t eq}, these conditions yield
\refstepcounter{equation}
$$
\label{transition scalings }
L_E=\mathcal{B}^{(2-5s_k)/9}, \quad T_E=\mathcal{B}^{(8-2s_k)/3}, \quad Q_E=\mathcal{B}^{10-s_k},\quad \Delta P_E = \mathcal{B}^{(20+4s_k)/9}.
\eqno{(\theequation{\mathrm{a}\!\!-\!\!\mathrm{d}})}
$$

Based on the above scalings, we introduce a strained coordinate centred at the edge of the uniform film,
\begin{equation}
\check{r}= \left(r - \mathcal{B}^{s_k}l_*\right)/\mathcal{B}^{(2-5s_k)/9}\nonumber  
\end{equation}
and pose the expansions
\begin{subequations}
 \label{transition expansions}
\begin{gather}
t= \mathcal{B}^{(8 - 2s_k)/3}\check{t}(\check{r})+\cdots, \quad q= \mathcal{B}^{10-s_k}\check{q}(\check{r})+\cdots,  \nonumber \tag{\theequation\!$\mathrm{a}$,$\mathrm{b}$}\\ 
p =2 + \mathcal{B}^2\hat{h}_{\infty} -\frac{\mathcal{B}^{2 + 2 s_k}}{2}l_*^2 + \mathcal{B}^{(20+4s_k)/9}\check{p}(\check{r})+\cdots.
\tag{\theequation\!c} 
\end{gather}
 \end{subequations}

The dynamic condition \eqref{p t eq} at leading order gives
\begin{equation}\label{p edge}
\check{p}=- l_* \check{r} - \frac{d^2 \check{t}}{d\check{r}^2}.
\end{equation}
Furthermore, substitution of \eqref{hat En} and (\ref{transition expansions}a,b) into the flux-conservation equation (\ref{q eqs}a) confirms that $\check{q}$ is uniform.  Matching with the uniform-film flux approximation (\ref{sol uniform}b) therefore gives
\begin{equation}\label{v}
\check{q} = \frac{1}{12}\left(6 \hat{\mathcal{E}}_k\right)^{3/4} l_*.
\end{equation}

Using (\ref{transition expansions}b)--\eqref{v}, the flux-pressure relation (\ref{q eqs}b) gives the differential equation
\begin{equation}\label{film transition eq}
\check{t}^3\left( l_*+\frac{d^3 \check{t}}{d\check{r}^3}\right)= \left(6 \hat{\mathcal{E}}_k\right)^{3/4} l_*
\end{equation}
governing the leading-order thickness profile. 
Matching with the uniform-film approximation (\ref{sol uniform}a) gives the far-field condition
\begin{equation}\label{w far field}
\check{t} \to  \left(6 \hat{\mathcal{E}}_k\right)^{1/4}\quad \text{as}\quad \check{r}\to-\infty.
\end{equation}
To close the problem governing $\check{t}$,  two additional far-field conditions are required in the diametric limit $\check{r}\to\infty$. 

The differential equation \eqref{film transition eq} has been studied by several authors, in particular in the  context of gravity-capillary waves at the interface of a two-dimensional thin film which flows at constant flux \cite{Wilson:83,Cuesta:12}. It has been shown that solutions of that equation which approach a constant value as $\check{r}\to-\infty$, as required by \eqref{w far field}, exhibit oscillations of growing amplitude as $\check{r}\to\infty$. Moreover, the latter oscillations are asymptotically comprised of an infinite chain of pairs of neck and bubble regions. Specifically, in the respective neck regions, \eqref{film transition eq} reduces to  
\begin{equation}
\check{t}^3 \frac{d^3 \check{t}}{d\check{r}^3}\approx \left(6 \hat{\mathcal{E}}_k\right)^{3/4} l_*,
\end{equation}
which we saw describes neck regions where vapour production is negligible (cf.~\eqref{dprimed eq}); furthermore, in the respective bubbles regions, \eqref{film transition eq} reduces to
\begin{equation}
\frac{d^3 \check{t}}{d\check{r}^3} \approx - l_*,
\end{equation}
which, upon integration, gives a radially localised version of the equations  we found to describe the unmodified bubble regions (cf.~\eqref{tilde t eq} and \eqref{bubeq}). It is therefore seen that the chain of neck-bubble pairs associated with the large $\check{r}$ behaviour of \eqref{film transition eq} constitutes a simplified version of the chain of neck-bubble pairs studied in \S\ref{sec:wave formation} and Appendix \ref{App:NBubbles}.  

While the oscillatory growth of $\check{t}$ forbids direct matching between the edge region and the $k$th-bubble region, the above observations suggest that it may be possible to match the latter bubble region with a formally infinite sequence of  edge-localised (i.e., two-dimensional) neck-bubble pairs, which, in turn, could be matched with the edge region,  similarly to the unconventional matching described in \S\ref{ssec:central}. We expect that such matching could be employed to close the edge-region problem \eqref{film transition eq} and \eqref{w far field}. 

The scalings of the edge-localised pairs can be obtained in a manner similar to that presented in \S\ref{sec:wave formation}. (Accordingly, the solution in each of these regions can be determined from an analysis similar to that carried out in Appendix \ref{App:NBubbles}). In particular, scalings for the outermost and largest edge-localised pair are shown in Fig.~\ref{fig:2seq}. We note that the chain of edge-localised pairs in the present limit evolved from pairs $m>k$ that existed in the distinguished limit \eqref{LimitInf}. Similarly, in the present limit we expect the $k$th pair to be continuously deformed from a non-localised to an edge-localised pair as $\mathcal{\hat{E}}_k$ decreases.  

\section{Expansion of uniform film into outermost bubble}\label{Sec:uniform2}
In this section, we carry out a more detailed analysis of the `last' distinguished limit, namely \eqref{Blimit2} for $k=1$. In that limit, the uniform film expands, as $\hat{\mathcal{E}}_1$ decreases,   into the outermost bubble region; all other neck-bubble pairs are localised to the edge of the uniform-film region, as described for arbitrary $k$ in the previous section. 

\begin{figure}[b!]
\begin{center}
\includegraphics[scale=0.295]{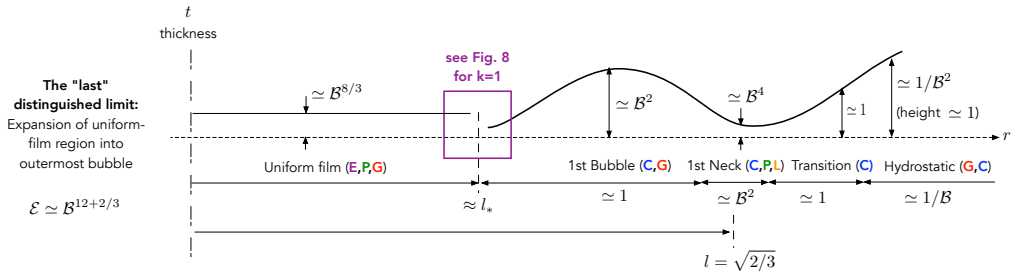}
\caption{Depiction of the `last' distinguished limit defined by \eqref{Blimit2} for $k=1$. The bold letters C, P, E, G and L have the same meaning as in Fig.~\ref{fig:lim1234}.}
\label{fig:last}
\end{center}
\end{figure}
The uniform-film region was already analysed in \S\ref{ssec:uniform 2}. In particular, the leading-order approximations \eqref{sol uniform} for the thickness, flux density and pressure-variation profiles in that region, respectively, hold for arbitrary $k$. In contrast, expression \eqref{l star} for the radial extent $l_*$ of the uniform film was derived assuming that the outermost neck-bubble pair remains unmodified by the expanding uniform film. Accordingly, it does not hold for the present case, $k=1$, where that pair is significantly modified by the expanding uniform film. In this section, we derive a revised description of that pair including a modified approximation for $l_*$.

\subsection{Outermost bubble}
The analysis of the outermost bubble region closely follows that of the bubble region in the second distinguished limit (\S\ref{ssec:second}), the main differences being the location and conditions satisfied at the inner boundary, as well as the negligibility of vapour production. (The analysis of the outermost bubble region in Appendix \ref{App:NBubbles}, where vapour production is negligible, is even closer.) In particular, the pressure, thickness and flux-density fields are expanded as in \eqref{Second bubble expansions}, where it is understood that the constant $\tilde{p}$ remains to be determined in the present limit. The leading-order thickness profile $\tilde{t}$ is governed by the differential equation \eqref{tilde t eq}. The bubble region is now bounded by the uniform-film edge $r=l_*$ and the outermost neck position $r=l>l_*$. (Since $p\sim 2$ in all regions inwards from the outermost neck, it follows from a leading-order balance of the force condition \eqref{force} that $l$ still equals $\sqrt{2/3}$.) At the inner boundary, indirect matching with the the uniform-film region, as discussed in \S\ref{ssec:edge} for arbitrary $k$, implies the conditions 
\refstepcounter{equation}
$$
\label{left bc last}
\tilde{t}=0, \quad \frac{d\tilde{t}}{dr}=0 \quad \text{at} \quad r=l_*.
\eqno{(\theequation{\mathrm{a},\!\mathrm{b}})}
$$
At the outer boundary, matching with the outermost neck implies the condition
\begin{equation}\label{right bc last}
\tilde{t}=0 \quad \text{at} \quad r=l.
\end{equation}

Solving \eqref{tilde t eq} together with \eqref{left bc last}, we find
\begin{equation}\label{last bubble t}
\tilde{t}=\frac{1}{32}\left(l_*^2-r^2\right)\left(r^2-\eta\right) + \frac{1}{16}{l_*}^2(l_*^2-\eta)\ln\frac{r}{l_*},
\end{equation}
where we have introduced the auxiliary parameter
\begin{equation}
\eta=8(\hat{h}_{\infty}-\tilde{p})-l_*^2,
\end{equation}
with $\hat{h}_{\infty}$ given by \eqref{Second varphi}. Applying the boundary condition \eqref{right bc last}, we obtain the relation
\begin{equation}\label{eta def}
\eta = l^2  + \left(l^2-l_*^2\right)\frac{2\ln\frac{l}{l_*}}{l^2/l_*^2-1 - 2\ln\frac{l}{l_*}}.
\end{equation}

Next, expansion of the flux-conservation equation (\ref{q eqs}a) in the present limit confirms that vapour production is negligible. The leading-order balance of that equation therefore reduces to a statement of continuity, whereby the leading-order flux density $\tilde{q}$ attenuates like $1/r$. In conjunction with matching with the uniform-film region, we find \begin{equation}\label{last bubble q}
\tilde{q}=\frac{l_*\hat{q}(l_*)}{r},
\end{equation}
wherein
\begin{equation}\label{hatq def}
\hat{q}(l_*) = \frac{l_*}{12}\left(6\hat{\mathcal{E}_1}\right)^{3/4}
\end{equation}
is the uniform-film approximation for the flux density (\ref{sol uniform}b) evaluated at the edge \eqref{uniform extent}.

The solutions \eqref{last bubble t} and \eqref{last bubble q} depend on the radial  extent $l_*$ of the uniform film. To determine this quantity we need to go beyond the matching condition \eqref{right bc last} and investigate the outermost neck in more detail.

\subsection{Outermost neck}
The analysis of the outermost neck region closely follows that of the neck region in the second distinguished limit (\S\ref{ssec:second}), the main difference being the negligibility of vapour production here. (The analysis of the outermost neck region in Appendix \ref{App:NBubbles}, where vapour production is negligible, is essentially identical except from the different matching conditions here owing to changes to the outermost bubble region.
In particular, we employ the strained coordinate $R$ defined by \eqref{R def}  and expand the thickness, flux density and pressure fields as in \eqref{Second neck expansions}. 

Expanding the flux-conservation equation (\ref{q eqs}a) in the present limit confirms that vapour production is negligible, as mentioned above. It readily follows that the leading-order flux density $Q(R)$ is constant, thence matching with the outermost bubble region gives
\begin{equation}\label{Q last limit def}
Q(R)= \tilde{q}(l),
\end{equation}
where $\tilde{q}(l)$ is obtained by evaluating \eqref{last bubble q} at $r=l$. 

Substitution of \eqref{Q last limit def} into a combined leading-order balance of the flux-pressure relation (\ref{q eqs}b) and dynamical condition \eqref{p t eq} yields  the differential equation
\begin{equation}\label{last T eq}
\frac{T^3}{12}\frac{d^3T}{dR^3}=\tilde{q}(l),
\end{equation}
which governs the leading-order thickness profile. 
Matching with the outermost bubble region, using \eqref{last bubble t}, yields the far-field condition
\begin{equation}
T\sim -\vartheta R \quad \text{as} \quad R\to-\infty,
\end{equation}
wherein
\begin{equation}\label{vartheta}
\vartheta =\frac{1}{16l}\left(l^2-{l_*}^2\right)\left(2l^2+l_*^2-\eta\right).
\end{equation}
Furthermore, matching with the transition region yields the far-field condition
\begin{equation}\label{last T far right}
 T \sim R^2 - 2 R_s R  
   \quad \text{as} \quad R\to\infty, 
\end{equation}
wherein $R_s$ corresponds to a constant shift of the neck radius. That constant was calculated in Appendix \ref{App:Force} assuming the second distinguished limit, giving the closed-form result \eqref{Rs text}. It is straightforward to revisit that derivation in the present limit to confirm that \eqref{Rs text} remains valid.
\begin{figure}[t!]
\begin{center}
\includegraphics[trim={2cm 0 0 0}, scale=0.29]{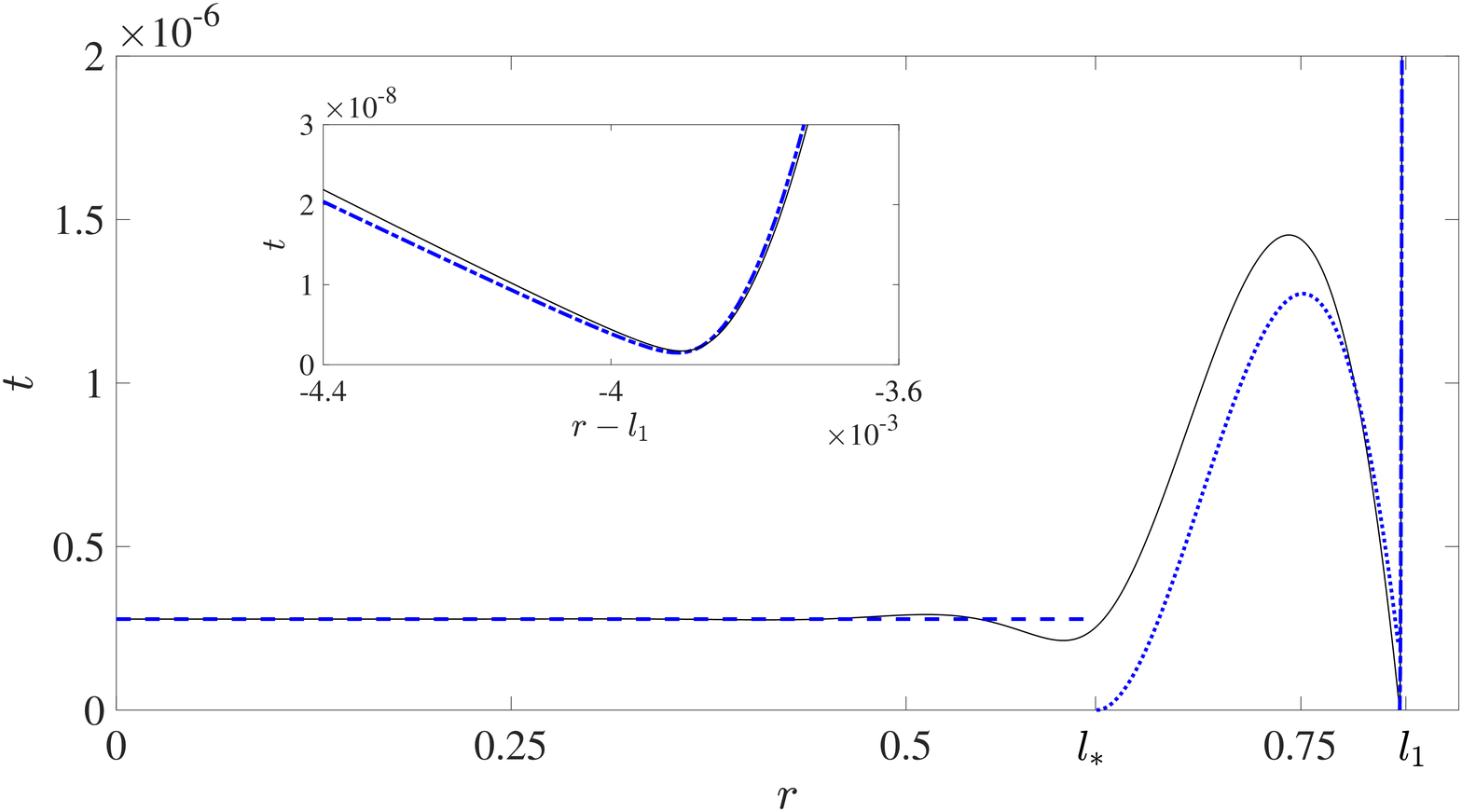}
\caption{Thickness profile for $\mathcal{B}=0.1$ and $\mathcal{E}=10^{-29}$. The solid curve depicts the numerical solution of the unified small-slope model (\S\ref{ssec:nondim}). The dashed (uniform-film), dotted (outermost bubble) and dash-dotted (outermost neck) curves depict the asymptotic approximations derived in \S\ref{Sec:uniform1} and \S\ref{Sec:uniform2} in the `last' distinguished limit, i.e., \eqref{Blimit2} with  $k=1$. Note the presence of localised capillary waves near the edge of the uniform-film region at $r=l_*$, as predicted in \S\ref{ssec:edge}. The inset focuses on the outermost-neck region near $r=l$.}
\label{llim}
\end{center}
\end{figure}

The neck problem consisting of \eqref{last T eq}--\eqref{last T far right} is of the form considered in Appendix \ref{App:CanNeckNum}. It follows that the constants $\vartheta$ and $\tilde{q}(l)$ are related as
\begin{equation} \label{theta q rel}
\vartheta=\left\{\tilde{q}(l)/\chi\right\}^{1/5},
\end{equation}
wherein $\chi$ is the numerically evaluated constant \eqref{chi value}. Combining \eqref{theta q rel} with \eqref{last bubble q}, \eqref{hatq def} and \eqref{vartheta}, we find the relation
\begin{equation}\label{last rel 2}
\left(l^2-l_*^2\right)\left(2l^2+l_*^2-\eta\right)=\frac{16 l_*^{2/5}l}{(12\chi l)^{1/5}}\left(6\hat{\mathcal{E}}_1\right)^{3/20}.
\end{equation}
Numerical solution of the pair of equations \eqref{eta def} into \eqref{last rel 2}  furnishes $l_*$ as a function of $\hat{\mathcal{E}}_1$. With $l_*$ determined, the leading-order asymptotic approximations describing the outermost bubble and neck regions are closed. 

In Fig.~\ref{llim}, we plot the thickness profile for $\mathcal{B}=0.1$ and $\mathcal{E}=10^{-29}$, with the solid line representing a numerical solution of the unified small-slope model and the dashed, dotted and dash-dotted lines representing the leading-order approximations obtained in the present limit for the uniform-film, outermost bubble and outermost neck regions, respectively. The uniform-film, bubble and neck regions are clearly visible in the numerical solution, as are the anticipated secondary oscillations at the edge of the uniform-film region. The particularly good agreement between the numerical solution and the uniform-film approximation suggests that the error in the latter is exponentially small (in contrast to the numerically appreciable algebraic errors in the bubble and neck regions). This can indeed be confirmed by higher-order scrutiny of the uniform-film  problem and the unconventional matching between that region and the outermost bubble region.

\subsection{Small-$\hat{\mathcal{E}}_1$ limit} 
From \eqref{eta def} and \eqref{last rel 2}, we find the behaviour
\begin{equation}
l_* - l \sim-\frac{\sqrt{6}\left(6\hat{\mathcal{E}}_1\right)^{3/40}}{(12\chi l^4)^{1/10}} \quad \text{as} \quad \hat{\mathcal{E}}_1\to0.
\end{equation}
This expression shows that the uniform film expands across the entire outermost bubble region. For sufficiently small $\hat{\mathcal{E}}_1$, the remainder of the outermost bubble region joins with the capillary waves at the edge of the uniform film. The resulting morphology is that of a uniform film bounded by capillary waves, as we observe numerically. We expect a similar morphology for moderately sized spheres, i.e., for $\mathcal{B}\simeq 1$, in the weak-evaporation limit, though this cannot be confirmed here as the unified small-slope lubrication model does not hold in that regime. In that sense, we expect the `last' limit considered in this section to be not just the last of the second sequence \eqref{Blimit2} of distinguished limits, but also the last limit relevant to the description of the evolution of the vapour film across the small-sphere regime. This is consistent with the fact, already noted in the introduction, that a similar morphology was found in the case of large drops levitated above a heated liquid bath \cite{Van:19}, although in that extreme scenario the gravity-dominated film region is not asymptotically uniform as it is here.

\section{Concluding remarks}\label{sec:discussion}

In this work, we used scaling arguments and matched asymptotics to study the levitation of a spherical particle above a liquid bath owing to the Leidenfrost effect. 
Motivated by numerical results for the closely related Leidenfrost scenario of a liquid drop levitated above a 
liquid bath \cite{Maquet:16}, we specifically set out to elucidate the evolution, with increasing sphere radius, of the vapour-film morphology. Little did we know, that evolution is asymptotically described by formally infinite sequences of distinguished limits, the limits being defined by the relative smallness of two dimensionless parameters: the Bond number and an intrinsic evaporation number.   
 
Let us briefly recapitulate the evolution of the film morphology, as revealed by our asymptotic analysis. Initially, i.e., for sufficiently small spheres, the evolution is similar to that in the classical Leidenfrost scenario of a liquid drop levitating above a flat substrate \cite{Celestini:12,Burton:12,Pomeau:12,Sobac:14}. Specifically, the thickness profile of the vapour film, as a function of the radial distance from the symmetry axis, transitions from a monotonic parabola to a non-monotonic profile, consisting of a relatively thick and wide bubble region bounded by a relatively thin and narrow neck region. While in the classical scenario this neck-bubble morphology is preserved as the drop size is increased further (until instability kicks in for sufficiently large drops \cite{Snoeijer:09,Quere:13}), in the present scenario the transition to a neck-bubble morphology merely constitutes the first stage of the evolution. The next stages are: (i) smaller and smaller neck-bubble pairs are sequentially formed near the origin; (ii) the pairs closest to the origin merge; (iii) the inner section of that merger transitions into a uniform-thickness film, which expands radially, gradually squeezing more and more neck-bubble pairs into a region of localised oscillations sandwiched between the uniform film and what remains of the bubble whose radial extent is presently comparable to the radial extent of the uniform film; and (iv) the uniform film expands across the original (largest and outermost) bubble region, the film morphology therefore reducing to a uniform film bounded by localised oscillations. See Figs.~\ref{fig:lim1234} and \ref{fig:1seq}--\ref{fig:last} for a schematic overview of these stages, including the scalings defining each of the distinguished limits as well as the regions participating in each of those limits. 

Assuming that the sphere density is comparable to that of the liquid bath, the entire evolution described above is found in the small-Bond-number regime corresponding to spheres that are small in comparison to the capillary length of the bath. Moreover, the ultimate morphology involving a nearly uniform film is essentially similar to that found asymptotically in \cite{Van:19} for drops much larger than the capillary length. This, as well as the numerical results presented in \cite{Maquet:16}, hints that further evolution of the film morphology beyond the stages reported here are likely quantitative rather than qualitative in nature. This provides \textit{a posteriori} reasoning for our focus on small Bond numbers.

Why, physically, is the evolution of the film morphology in the present problem so much more complicated than in the classical Leidenfrost scenario? 
A key difference has to do with the manner in which gravity affects the vapour film. In the classical scenario, variations in gravitational-hydrostatic pressure begin to affect the bubble region when the drop size becomes comparable to the capillary length \cite{Sobac:14}. In contrast, in the present scenario the bubble region is affected by such variations already at small Bond numbers; this is because height variations across the bubble region are not just due to film thickness but also to the fact that the film centreline follows the curved boundary of the sphere. These pressure variations result in the breakdown of the first distinguished limit considered in our analysis, which is analogous to the small-Bond limit in the classical scenario; this asymptotic breakdown, first of many, therefore marks the sphere-radius scaling above which the present problem significantly departs from the classical one. We emphasise that this comparison is made with the classical scenario assuming a flat solid substrate; in the case where the solid substrate is curved, as is often the case in experiments, the above discussion actually suggests an intricate evolution similar to the one in the present problem.  Conversely, analyses of drops levitated above a curved substrate (owing to uniform injection of gas through the substrate rather than the Leidenfrost effect) suggest that multiple static solutions may exist for non-small Bond numbers \cite{Duchemin:05,Lister:08}.

Following previous theoretical studies of the Leidenfrost effect, we carried out our asymptotic analysis starting from a reduced formulation based on a lubrication approximation. Specifically, we derived a `unified small-slope model' to simultaneously describe both the vapour film underneath the sphere and the hydrostatic region of the liquid bath away from the sphere. For spheres whose density is comparable to that of the bath, it can be readily seen that the small-Bond-number approximations we have derived based on the unified small-slope model are indeed consistent with the assumptions used to derive that model (unless evaporation is too strong, see remark 4 in  \S\ref{ssec:hydro}). In particular, a key assumption was that the slope of the liquid-bath height profile is everywhere small. This assumption is expected to fail, e.g., at moderate Bond numbers, or even small Bond numbers in the case of heavy spheres whose density greatly exceeds that of the liquid bath. In those cases, and assuming evaporation is sufficiently weak, an adequate lubrication model could still be devised, as done in \cite{Van:19} for drops levitated above a bath. In that paper, the lubrication model is based on a suitable parameterisation of the film's centreline, rather than the use of cylindrical coordinates; moreover, the exact nonlinear form of the curvature of the bath interface is used, rather than the linearised form used herein. 

We conclude by proposing several extensions to this work. An obvious one is to consider a drop, rather than spherical particle, levitated above a liquid bath. For small Bond numbers, it can be shown that the drop scenario can be mathematically mapped to the present one. The mapping is not straightforward yet serves to show that the evolution of the film morphology in the drop scenario is essentially similar to that found here in the spherical-particle scenario. (A similar mapping has been used in \cite{Yiantsios:90} in the analysis of a small drop approaching a liquid bath, though gravitational-hydrostatic contributions were neglected in that paper.) A less direct extension would be to study the evolution of the film morphology for heavy spherical particles, or drops, in which case we expect the key stages of the evolution of the film morphology to occur in a small-Bond-number regime where the centreline of the film is significantly curved; this would require employing a generalised lubrication model along the lines discussed in the preceding paragraph. 

It is also desirable to extend our modelling to include the dynamical effects of the bath flow, which we have neglected from the start. The latter approximation is typical in analytical studies of the Leidenfrost effect and can be theoretically justified for sufficiently viscous liquids. Recent experiments and numerical calculations have shown, however, that bath flows driven by vapour-shear and Marangoni stresses can significantly affect the Leidenfrost levitation of a drop above a bath \cite{Sobac:20}. Including these effects would entail coupling the vapour-film and liquid-bath flows. In particular, for small-Bond numbers and moderately dense particles, where the bath is nearly flat, the bath domain could perhaps be approximated by a half-space.

Last, the modelling could be extended to dynamical scenarios in which the levitated spherical particle is non-stationary. This would require coupling the equations of motion of the particle with a modified lubrication model from which the hydrodynamic loads on the particle can be calculated given the particle's linear and angular velocities \cite{Brandao:20}. In particular, it would be of interest to apply such modelling to illuminate recent experiments showing that inverse-Leidenfrost spherical particles and drops levitated above an evaporating cryogenic liquid bath can move spontaneously in the absence of external forces \cite{Gauthier:19}. 

\textbf{Acknowledgements.} O.~Schnitzer is grateful for the financial support from the Leverhulme Trust through Research Project Grant RPG-2021-161.

\appendix

\section{Higher-order force balance}\label{App:Force}
In this appendix we calculate the constant shift $R_s$, which corresponds to an order $\mathcal{B}^2$ shift of the neck profile (see \S\ref{ssec:second}). To this end, we shall consider a higher-order balance of the force condition \eqref{force}. Thus, consider that condition in the form
\begin{equation}\label{p decom}
\int_0^{l+\delta}pr\,dr+\int_{l+\delta}^\infty pr\,dr=\frac{2}{3},
\end{equation}
in which $\delta$ is a splitting parameter chosen such that 
\begin{equation}\label{res1}
\mathcal{B}^2\ll \delta \ll \mathcal{B}.
\end{equation}

The second integral in \eqref{p decom} is $o(\mathcal{B}^2)$. To see this, note that the pressure in the transition region  $\simeq\mathcal{B}^{10}$; moreover, from considerations of asymptotic matching with the neck region, the leading-order pressure field in the transition region $\simeq1/(r-l)^5$ as $r\searrow l$, up to logarithmic factors. This suggests a dominant contribution to the integral $\simeq\mathcal{B}^{10}/\delta^4$, which is clearly $\ll \mathcal{B}^2$.

Next, we use the dynamic condition \eqref{p t eq}, the symmetry condition (\ref{bcs at zero}a) together with \eqref{t def}, and the restriction \eqref{res1} to  write the first integral in \eqref{p decom} as
\begin{equation}\label{first int}
\int_0^{l+\delta}pr\,dr=  l^2 + 2 \delta l + \mathcal{B}^2\left(\hat{h}_{\infty}-\frac{l^2}{4}\right)\frac{l^2}{2} + \mathcal{B}^2\int_0^{l + \delta}  r t \, dr - (l  + \delta ) \frac{d t}{dr}\bigg|_{r = l + \delta} +o(\mathcal{B}^2),
\end{equation}
where, consistently with the indicated error, we have approximated the sphere displacement $h_{\infty}$ by its leading-order value $\hat{h}_{\infty}$ (cf.~\eqref{phi far}). 

The integral term on the right-hand side of \eqref{first int} is $o(\mathcal{B}^2)$ and so can be incorporated in the error term. To see this, recall that $t$ is asymptotically small in both the bubble and neck regions, whereas $r\simeq 1$ in those regions; furthermore, the leading-order approximation $\mathcal{B}^4T(R)$ for the thickness profile in the neck region grows quadratically in terms of the neck coordinate $R=(r-l)/\mathcal{B}^2$. The integral term is therefore estimated to be $O(\delta^3)$, which using \eqref{res1} is $o(\mathcal{B}^2)$. 

Next, approximating the derivative term in \eqref{first int} based on the leading-order approximation for $t$ in the neck region gives
\begin{equation}\label{first dev}
\frac{d t}{dr}\bigg|_{r = l + \delta}\approx \mathcal{B}^2\frac{d T}{dR}\bigg|_{R = \delta/\mathcal{B}^2}={\tilde{\mathcal{E}}}^{1/6}\mathcal{B}^2\frac{d T'}{dR'}\bigg|_{R'= \frac{\delta/\mathcal{B}^2-R_s}{\tilde{\mathcal{E}}^{1/6}}}\\=2 \delta -2R_s\mathcal{B}^2+o(\mathcal{B}^2),
\end{equation} 
where in the second step we used the  transformations \eqref{sim trans} and \eqref{Rs def}, while in the last step we used the far-field expansion $T'=R'^2+O(1)$ as $R'\to\infty$ (cf.~(\ref{Second neck local right}a)). 

Combining the above results into \eqref{p decom}, and using $l=\sqrt{2/3}$, we find from the order $\mathcal{B}^2$ balance of that equation that 
\begin{equation}\label{Rs app}
R_s=\left(l^2-4 \hat{h}_{\infty}\right) \frac{l}{16}.
\end{equation}
The dependence upon the arbitrary splitting parameter $\delta$ has vanished, as it must. 

We remark that in deriving \eqref{first int} and \eqref{first dev} we tacitly ignored higher-order corrections to the thickness profile in the neck region. It can be shown that these corrections grow at most quadratically with $R$ (up to terms of very high order in $\mathcal{B}$), which immediately justifies them being discarded in \eqref{first int}. More care is due in \eqref{first dev}, however, where a correction to the neck thickness profile at order $\mathcal{B}^4\lambda(\mathcal{B})$, say, where $\lambda(\mathcal{B})\ll1$, gives an $O(\lambda\delta)$ contribution; this is clearly small compared to the leading term in \eqref{first dev}, but not necessarily $o(\mathcal{B}^2)$ unless $\lambda=o(\mathcal{B})$. Otherwise, the error can be made $o(\mathcal{B}^2)$ by simply refining the restriction \eqref{res1} to $\mathcal{B}^2\ll\delta\ll\mathcal{B}^2/\lambda$. 

\section{Generalised constant-flux neck problem}\label{App:CanNeckNum}

Consider the third-order differential equation
\begin{equation}\label{TT ODE}
\frac{\mathcal{T}^3}{12}\frac{d^3\mathcal{T}}{dX^3} = \mathcal{Q},
\end{equation}
 subject to the far-field conditions
\refstepcounter{equation}
$$
\label{TT far}
\mathcal{T}\sim -\theta_{-} X \quad \text{as} \quad X\to-\infty,\quad\mathcal{T}\sim  \kappa X^2 + \theta_{+} X + O(1) \quad \text{as} \quad X\to\infty,
\eqno{(\rm{\theequation}{\mathrm{a},\!\mathrm{b}})}
$$
wherein $\theta_{-},\theta_{+},\kappa$ and $\mathcal{Q}$ are constants.

Employing the transformations 
\refstepcounter{equation}
$$
X= \frac{\theta_-}{\kappa} \bar{X} -  \frac{\theta_+}{2 \kappa}, \quad \mathcal{T} = \frac{\theta_-^2}{\kappa} \bar{\mathcal{T}},
\eqno{(\rm{\theequation}{\mathrm{a},\!\mathrm{b}})}
$$
the differential equation \eqref{TT ODE} becomes
\begin{equation}
\frac{\bar{\mathcal{T}}^3}{12}\frac{d^3\bar{\mathcal{T}}}{dX^3} = \mathcal{Q}\kappa\theta_{-}^{-5},
\end{equation}
while the far-field conditions \eqref{TT far} become
\refstepcounter{equation}
$$
\mathcal{\bar{T}}\sim -\bar{X} \quad \text{as} \quad \bar{X}\to-\infty , \quad \mathcal{\bar{T}} = \bar{X}^2 + O(1) \quad \text{as} \quad \bar{X}\to\infty.
\eqno{(\rm{\theequation}{\mathrm{a},\!\mathrm{b}})}
$$

The transformed problem is seen to be identical to the constant-flux neck problem \eqref{dprimed eq} and \eqref{dprimed far fields}, with $\chi=\mathcal{Q}\kappa \theta_{-}^{-5}$. As further discussed in \S\ref{ssec:limitneck}, numerical solution of the latter problem yields $\chi\doteq  0.051$.

\section{Mapping between sub-film and thin-film problems}\label{App:ThirdBreak}
In this appendix we relate the sub-film problem formulated in \S\ref{sssec:subfilm} with the thin-film problem formulated in \S\ref{ssec:thinfilm}. Considering the sub-film problem, the far-field behaviour (\ref{check far}a) can be extended as
\begin{equation} 
\check{t}\sim \frac{1}{48}\check{r}^2-c\ln \check{r} \quad \text{as} \quad \check{r}\to\infty,
\end{equation}
wherein the constant coefficient $c$ is an output of that problem. With this extension, the sub-film problem can be mapped to the thin-film problem upon making the transformations
\refstepcounter{equation}
$$
\label{appendix transformations}
\check{r}=6 c^{1/2} \check{r}', \quad \check{t}=\frac{3c}{2}\check{t}', \quad \check{p}=\hat{h}_{\infty}+\frac{1}{24}(\check{p}'-2), \quad \check{q}=\frac{c^{5/2}}{128}\check{q}', \quad \tilde{\mathcal{E}}=\frac{3c^3}{512}\tilde{\mathcal{E}}',
\eqno{(\rm{\theequation}{\mathrm{a}\!\!-\!\!\mathrm{e}})}
$$
with $\check{r}',\check{t}',\check{p}',\check{q}'$ and $\tilde{\mathcal{E}}'$ substituted instead of $r,t_0,p_0,q_0$ and $\mathcal{E}$, respectively. To complete the linkage between the two problems, it remains to determine $c$. Thus, on the one hand, the thin-film problem implies the far-field behaviour 
\begin{equation}
\check{q}'\sim 2\tilde{\mathcal{E}}'\frac{1}{\check{r}'}\ln\check{r}' + \frac{\kappa}{\check{r}'} \quad \text{as} \quad \check{r}'\to\infty,
\end{equation}
where $\kappa$ is a function of $\tilde{\mathcal{E}}'$ determined as part of that problem. (This follows from local analysis of (\ref{First eqs}a) using \eqref{First t def} and \eqref{First h far}.) On the other hand, the sub-film condition includes the far-field condition (\ref{check far}b), which here gives
\begin{equation}\label{our kappa}
\kappa/\tilde{\mathcal{E}}'=2\ln (6c^{1/2}) +l^2\ln \mathcal{B}+\zeta,
\end{equation}
in which $\zeta$ is defined by \eqref{zeta def} as a function of $\tilde{\mathcal{E}}$ and $\ln\mathcal{B}$. This comparison thus furnishes a relation from which $c$ could in principle be calculated as a function of $\mathcal{E}$ and $\ln\mathcal{B}$. 

Consider now the limit $\tilde{\mathcal{E}}\to0$ of the sub-film problem. Specifically, for $\tilde{\mathcal{E}}\ll 1/\ln\mathcal{B}$, substituting (\ref{tau nu asym}a) and \eqref{Third theta} into \eqref{zeta def} gives $\zeta\simeq 1/\tilde{\mathcal{E}}$. Assuming  $1/\tilde{\mathcal{E}}\gg \ln c$,  we then have from (\ref{appendix transformations}e) and \eqref{our kappa} that $\kappa \simeq 1/c^3$. Furthermore, it can be shown from the asymptotics of the thin-film problem that $\kappa\simeq \mathcal{E}'^{5/6}$. (While the latter asymptotics are not explicitly provided in the text, they are easily extracted  from the analysis of the second distinguished limit.) Comparing the latter two estimates for $c$ and using (\ref{appendix transformations}e), we find $
c\simeq \tilde{\mathcal{E}}^{-5/3}$, consistently with our assumption above, and $\tilde{\mathcal{E}}'\simeq \tilde{\mathcal{E}}^6$.  The latter estimate shows  that the behaviour as $\tilde{\mathcal{E}}\to0$ of the solution to the  sub-film problem is analogous to the behaviour as $\mathcal{E}\to0$ of the solution to the thin-film problem. This observation is used in \S\ref{ssec:thirdBD} to infer the breakdown of the third distinguished limit. 

\section{Analysis for $n$ neck-pocket pairs}\label{App:NBubbles}
We here analyse the distinguished limit corresponding to the $n$th element of the sequence \eqref{Blimit}, with $n=2,3,\ldots$ (the case $n=1$ corresponds to the second distinguished limit studied in \S\ref{ssec:second}). To this end, we define the corresponding rescaled evaporation number
\begin{equation}
\mathcal{E}_n=\mathcal{E}/\mathcal{B}^{12+2s_n},
\end{equation}
where $s_n$ is given by \eqref{sm}. As discussed in \S\ref{ssec:seqlimits}, the present  limit corresponds to a scenario where the solution domain decomposes into a chain of $n$ neck-bubble pairs in addition to the transition and hydrostatic regions. These are enumerated as in \S\ref{ssec:scaling analysis} by the index $m=1,2,\ldots,n$, with $m=1$ corresponding to the outermost pair, whose associated neck region matches to the transition region, and $m=n$ to the innermost pair, with the $n$th bubble centred about the origin. 

The analysis in this appendix builds on the scaling analysis in \S\ref{ssec:scaling analysis}. In particular, that analysis implies that vapour production is only non-negligible in the innermost pair. We shall therefore  separately analyse the outer pairs $m<n$ and the innermost pair $m=n$.

\subsection{Outer bubble regions  ($m<n$)}
Given the scalings  (\ref{scalings sm}a--c), to analyse the $m$th bubble region we introduce the strained coordinate $r_m = r/\mathcal{B}^{s_m}$ and expand the thickness profile and flux density as 
\refstepcounter{equation}
$$
\label{tq app}
t= \mathcal{B}^{2+4s_m}\tilde{t}_m(r_m) + \ldots, \quad q= \mathcal{B}^{10-s_m}\tilde{q}_m(r_m)+ \ldots,
\eqno{(\rm{\theequation}{\mathrm{a},\!\mathrm{b}})}
$$
respectively, and the pressure field as (cf.~\eqref{P expansion bubble})
 \begin{subequations}
\label{mpbub}
\begin{gather}
p= 2 + \mathcal{B}^2\tilde{p}_1 + \cdots, \quad \text{for} \quad m=1,  \nonumber \tag{\theequation\!a} \\
p= 2 + \mathcal{B}^2\hat{h}_{\infty}+\mathcal{B}^{2+2s_m}\tilde{p}_m + \cdots, \quad \text{for} \quad m=2,3,\ldots,n, \nonumber \tag{\theequation\!b} 
\end{gather} 
\end{subequations}
where $\hat{h}_{\infty}$ is given by \eqref{varphi} and $\tilde{p}_m$ are constants. The domain of the $m$th bubble region is $0<r_m<l_m$, where the scaled radial extent of the $m$th bubble region is denoted $l_m$, consistently with the notation used below \eqref{P expansion neck}. The constants $\tilde{p}_m$ and $l_m$ remain to be determined. 

In this subsection we consider the outer bubble regions $m<n$. For those bubble regions, the flux-conservation equation (\ref{q eqs}a) confirms that vapour production is negligible and thence that
\begin{equation}
\label{flux bub app}
\tilde{q}_m = \frac{j_m}{r_m},
\end{equation}
where $j_m$ are constants to be determined. Next, by expanding the dynamic condition \eqref{p t eq} we arrive at an equation for the thickness profile,
\begin{equation}\label{bubeq}
\frac{1}{r_m}\frac{d}{dr_m}\left(r_m\frac{d\tilde{t}_m}{dr_m}\right)= \delta_{m1}\hat{h}_{\infty}-\frac{1}{2}r_n^2 - \tilde{p}_m,
\end{equation}
where $\delta_{ij}$ is the Kronecker delta symbol.

As discussed in \S\ref{ssec:scaling analysis}, matching with the neck regions demands that the thickness profile $\tilde{t}_m$ vanishes quadratically at $r_m=0$ and linearly at $r_m=l_m$. It follows that the thickness profile satisfies the boundary conditions 
\refstepcounter{equation}
$$
\label{bubbc}
t_m = 0,\quad \frac{d\tilde{t}_m}{dr_m}=0 \quad \text{at} \quad r_m=0, \quad \tilde{t}_m=0 \quad \text{at} \quad r_m=l_m,
\eqno{(\rm{\theequation}{\mathrm{a},\!\mathrm{b}})}
$$
Solving \eqref{bubeq}--\eqref{bubbc}, we find
\begin{equation}\label{bprofm}
\tilde{t}_m=\frac{1}{32}(l_m^2-r_m^2)r_m^2
\end{equation}
together with the relation
\begin{equation}\label{pl}
\tilde{p}_m=\delta_{m,1}\hat{h}_{\infty}-\frac{l_m^2}{8}.
\end{equation}

A leading-order balance of the force condition \eqref{force} shows that 
\begin{equation}\label{l1 is l}
l_1=l, 
\end{equation}
where $l=\sqrt{2/3}$ as in the main text. The corresponding pressure correction $\tilde{p}_1$ then follows from \eqref{pl}. The remaining extents and corresponding pressure corrections remain to be determined. 

\subsection{Outer neck regions ($m<n$)}
We next turn to the neck regions. Given the scalings  (\ref{scalings sm}d--f), to analyse the $m$th neck region we introduce the strained coordinate
\begin{equation}
R_m=(r-l_m\mathcal{B}^{s_m})/ \mathcal{B}^{4 - 10 s_m}
\end{equation}
and
expand the thickness and flux density as
\refstepcounter{equation}
$$
\label{tq neck app}
T= \mathcal{B}^{4-10s_m}T_m(R_m) + \ldots, \quad q= \mathcal{B}^{10-s_m}Q_m(R_m)+ \ldots,
\eqno{(\rm{\theequation}{\mathrm{a},\!\mathrm{b}})}
$$
respectively, and the pressure as (cf.~\eqref{P expansion neck})
 \begin{subequations}
  \label{mpneck}
 \begin{gather}
p= P(R_1) + \cdots \quad \text{for} \quad m=1, \nonumber \tag{\theequation\!a} \\
p= 2+\mathcal{B}^2P_2(R_2)+ \cdots , \quad \text{for} \quad m=2,  \nonumber \tag{\theequation\!b} \\
p= 2+\mathcal{B}^2\hat{h}_{\infty}+\mathcal{B}^{16s_m}P_m(R_m)+ \cdots , \quad \text{for} \quad m=3,4,\ldots,n,  \nonumber \tag{\theequation\!c} 
\end{gather} 
\end{subequations}
where $\hat{h}_{\infty}$ is given by \eqref{varphi}. 

We here consider the outer neck regions $m<n$. For those neck regions, the flux-conservation equation (\ref{q eqs}a) confirms that vapour production is negligible and implies a uniform flux density at leading order, say 
\begin{equation}\label{NB flux const}
Q_m(R_m)\equiv \tilde{Q}_m.
\end{equation}
Since vapour production is, in fact, negligible for all outer pairs of neck and bubble regions, continuity implies 
\refstepcounter{equation}
$$
\label{Qrelapp}
j_m=Q_1 l, \quad \tilde{Q}_m l_m= \tilde{Q}_1 l. 
\eqno{(\rm{\theequation}{\mathrm{a},\!\mathrm{b}})}
$$
These relations can be confirmed as follows. First, for $m=1,2,\ldots,n-1$, matching the flux density between the $m$th neck and bubble regions gives $\tilde{Q}_{m}l_m= j_m$. Second, for $m=2,\ldots,n-1$, matching the flux density between the $m$th neck region and the $(m-1)$th bubble region gives $\tilde{Q}_{m}l_m=j_{m-1}$. Combining these relations and using  \eqref{l1 is l} gives \eqref{Qrelapp}. 

With \eqref{NB flux const}, the flux-pressure relation (\ref{q eqs}b) and dynamic condition \eqref{p t eq}  give
\refstepcounter{equation}
$$
 \tilde{Q}_m=-\frac{T_m^3}{12}\frac{dP_m}{dR_m}, \quad P_m=2\delta_{m,1}+\hat{h}_{\infty}\delta_{m,2}-\frac{d^2T_m}{dR_m^2},
\eqno{(\rm{\theequation}{\mathrm{a},\!\mathrm{b}})}
$$
which can be combined into an equation for the thickness profile,
\begin{equation}\label{mneck}
\frac{T_m^3}{12}\frac{d^3T_m}{dR_m^3}= \tilde{Q}_m.
\end{equation}

Matching the thickness between the $m$th neck and its two neighbouring bubble regions yields the far-field conditions (cf.~\eqref{bprofm})
\refstepcounter{equation}
$$
\label{NB far}
T_m \sim -\sigma_m R_m\quad \text{as} \quad R_m\to-\infty, \quad T_m \sim \kappa_m R_m^2 \quad \text{as} \quad R_m\to\infty,
\eqno{(\rm{\theequation}{\mathrm{a},\!\mathrm{b}})}
$$
where
\begin{equation}\label{thetam}
\sigma_{m} = l_m^3/16
\end{equation}
and
\refstepcounter{equation}
$$
\label{km}
\kappa_1=1, \quad \kappa_m=\frac{l_{m-1}^2}{16} \quad \text{for} \quad m=2,3,\ldots n-1.
\eqno{(\rm{\theequation}{\mathrm{a},\!\mathrm{b}})}
$$
(Note that the outer necks match only with the outer bubbles which we have analysed in the preceding subsection.) 

For each $m$, the differential equation \eqref{mneck} and far-field conditions \eqref{NB far} together constitute a constant-flux neck problem of the type formulated in Appendix \ref{App:CanNeckNum}. Using the main result in that appendix, we obtain the relation 
\begin{equation}\label{NBchirel}
\chi \sigma_m^5 = \tilde{Q}_m \kappa_m.
\end{equation}

\subsection{Recursive solution for the outer neck-bubble pairs ($m<n$)}
For $m=1$, combining \eqref{l1 is l} with  \eqref{thetam}--\eqref{NBchirel} gives 
\begin{equation}\label{Q1app}
 \tilde{Q}_1 = \frac{\chi}{(12\sqrt{6})^5}.
\end{equation}
Then, using \eqref{thetam} and \eqref{km} for two subsequent neck regions, with the corresponding flux densities related through \eqref{Qrelapp} (using \eqref{Q1app}), 
we find the recurrence relation
\begin{equation}\label{NP rec}
l_{m}=\left(\frac{1}{2^{1/4}}\sqrt{\frac{2}{3}}\right)l_{m-1}^{1/8} \quad \text{for} \quad m=2,3,\ldots,n-1. 
\end{equation}
Solving \eqref{NP rec} with the initial condition \eqref{l1 is l}, we find 
\begin{equation}\label{NB lm}
l_m=\left(\frac{1}{2^{1/4}}\sqrt{\frac{2}{3}}\right)^{8 s_{m}}\left(\frac{2}{3}\right)^{4/8^m}\quad \text{for} \quad m=1,2,\ldots,n-1.
\end{equation}
The respective values of $\tilde{Q}_m$  can be obtained from \eqref{NBchirel} using \eqref{thetam}, \eqref{km} and \eqref{NB lm}. 

The above results fully  determine the leading-order approximations for the outer bubble regions. Leading-order solutions for the outer neck regions are determined by the constant-flux neck problems \eqref{mneck}--\eqref{km} up to constant radial shifts on the order of the respective neck widths. (The shift of the outermost neck can be shown to be the same as that calculated in Appendix \ref{App:Force} assuming the second distinguished limit.)  
Note that we have determined these leading-order solutions without yet considering the innermost neck-bubble pair. In particular, we see that the approximation
\begin{equation}\label{cont flux app}
q \sim \mathcal{B}^{10} \frac{\chi\sqrt{2/3}}{(12\sqrt{6})^5}\frac{1}{r}
\end{equation}
holds across all of the outer pairs, regardless of the number of pairs $n$. 

\subsection{Innermost neck-bubble pair ($m=n$)}
We now consider the innermost bubble region. 
The thickness profile of the bubble still satisfies the differential equation \eqref{bubeq}. In the present case, it is to be solved in conjunction with the symmetry condition (cf.~(\ref{bcs at zero}a))
\begin{equation}
\frac{d\tilde{t}_n}{dr_n}=0 \quad \text{at} \quad r_n=0,
\end{equation}
and the boundary condition 
\begin{equation}
\quad \tilde{t}_n=0 \quad \text{at} \quad r_n=l_n,
\end{equation}
which follows from matching with the $n$th neck region. The solution is found as
\begin{equation}\label{tn vap app}
\tilde{t}_n=\frac{1}{32}(l_n^2-r_n^2)(\alpha_n+r_n^2),
\end{equation}
wherein
\begin{equation}
\alpha_n=l_n^2+8 \tilde{p}_n .
\end{equation}
Note that $l_n$ and $\tilde{p}_n$, and hence $\alpha_n$, remain to be determined. 

Vapour production is non-negligible for the innermost neck-bubble pair, in contrast to what we have seen for the outer pairs. Thus, for the innermost bubble region the flux-conservation equation (\ref{q eqs}a) gives 
\begin{equation}
\label{flux n app}
\frac{1}{r_n}\frac{d}{dr_n}\left(r_n\tilde{q}_n\right)=\frac{\mathcal{E}_n}{\tilde{t}_n},
\end{equation}
to be solved together with the symmetry condition (cf.~(\ref{bcs at zero}b))
\begin{equation}
\label{flux bc app}
\tilde{q}_n=0 \quad \text{at} \quad r_n=0.
\end{equation}
Integration of \eqref{flux n app} using \eqref{tn vap app} and \eqref{flux bc app} gives
\begin{equation}\label{NB qn}
\tilde{q}_n=\frac{16\mathcal{E}_n}{r_n(l_n^2+\alpha_n)}\ln\frac{l_n^2(r_n^2+\alpha_n)}{\alpha_n(l_n^2-r_n^2)}.
\end{equation}

Consider now the innermost neck region. From the lubrication equations  \eqref{q eqs} and the dynamic condition \eqref{p t eq}, we find 
\refstepcounter{equation}
$$
\label{NB neck eqs}
\frac{dQ_n}{dR_n}=\frac{\mathcal{E}_n}{T_n}, \quad Q_n=-\frac{T_n^3}{12}\frac{dP_n}{dR_n}, \quad P_n=-\frac{d^2T_n}{dR_n^2}.
\eqno{(\rm{\theequation}{\mathrm{a}\!\!-\!\!\mathrm{c}})}
$$
Matching thickness between the $n$th neck and the adjacent bubble regions furnishes the far-field conditions (cf.~\eqref{bprofm} and \eqref{tn vap app})
\refstepcounter{equation}
$$
\label{NB neck far1}
T_n\sim-\varsigma_n R_n\quad \text{as} \quad R_n\to-\infty, \quad T_n\to \kappa_{n} R_n^2 \quad \text{as} \quad R_n\to\infty,
\eqno{(\rm{\theequation}{\mathrm{a},\!\mathrm{b}})}
$$
where 
 \begin{equation}
\varsigma_n =\frac{l_n(l_n^2+\alpha_n)}{16}
\end{equation}
and $\kappa_n$ is given by \eqref{km}. Similarly matching the flux density gives (cf.~\eqref{flux bub app}, (\ref{Qrelapp}a) and \eqref{NB qn}) the far-field conditions 
\refstepcounter{equation}
$$
\label{NB neck far2}
 Q_n\sim -\frac{1}{\varsigma_n}\ln \frac{\varsigma_n |R_n|}{2\alpha_n\mathcal{B}^{2-14s_n}} \ \text{as} \ R _n\to-\infty, \quad Q_n\sim  \frac{\chi\sqrt{2/3}}{(12\sqrt{6})^5}\frac{1}{l_{n}} \ \text{as} \  R_n\to\infty.
\eqno{(\rm{\theequation}{\mathrm{a},\!\mathrm{b}})}
$$

The neck problem consisting of the differential equations \eqref{NB neck eqs} and far-field conditions \eqref{NB neck far1} and \eqref{NB neck far2} is similar to the neck problem in the second distinguished limit (see \S\ref{sssec:necksecond}). Following the same solution procedure, we can numerically solve for the leading-order thickness profile of the innermost neck region (up to a small constant radial shift on the order of the neck width), along with the constants $l_n$, $\alpha_n$ and $\tilde{p}_n$ as functions of $\mathcal{E}_n$ and $\ln\mathcal{B}$. The latter constants then fully determine the leading-order thickness profile of the innermost bubble region (cf.~\eqref{tn vap app}). 

\bibliography{refs}
\end{document}